\documentclass[
	aps,physrev,reprint,a4paper,
	superscriptaddress,nofootinbib,
	amsmath,amssymb,floatfix,
	nobalancelastpage, 
	raggedbottom 
]{revtex4-2}

\usepackage[T1]{fontenc}
\usepackage{lmodern}
\usepackage{graphicx}
\usepackage{mathtools}
\usepackage{physics}
\usepackage{bm}
\usepackage{microtype}
\usepackage{xcolor}

\usepackage[pdfusetitle]{hyperref}
\hypersetup{
	unicode=true,
	breaklinks=true,
	bookmarksnumbered=true,
	bookmarksopen=true,
	colorlinks=true,
	linkcolor=blue,
	citecolor=blue,
	urlcolor=blue,
	pdfauthor={Urei Miura and Hisao Hayakawa},
	pdfsubject={},
	pdfkeywords={}
}

\usepackage{tikz}
\usepackage{pgfplots}
\usetikzlibrary{arrows.meta,decorations.markings}
\usepgfplotslibrary{groupplots}
\pgfplotsset{compat=1.18}
\definecolor{paperblue}{HTML}{0072B2}
\definecolor{paperorange}{HTML}{D55E00}
\definecolor{paperteal}{HTML}{009E73}
\definecolor{paperpurple}{HTML}{7A5195}
\definecolor{paperink}{HTML}{24292E}
\definecolor{paperguide}{HTML}{8A949E}
\definecolor{paperfaint}{HTML}{EDF1F4}
\newcommand{\PaperLabelFont}{\fontsize{8.5}{10.2}\selectfont}
\newcommand{\PaperSmallFont}{\fontsize{7.5}{9}\selectfont}
\newcommand{\PaperPanelFont}{\fontsize{9.5}{11.4}\selectfont\bfseries}
\tikzset{
	paper axis/.style={draw=paperink,line width=.55pt},
	paper guide/.style={draw=paperguide,line width=.45pt},
	paper curve/.style={line width=1.15pt,line cap=round,line join=round},
	paper panel/.style={font=\PaperPanelFont,anchor=north west,inner sep=0pt},
	paper label/.style={font=\PaperLabelFont,inner sep=1.5pt}
}

\newcommand{\C}{\mathbb{C}}
\newcommand{\R}{\mathbb{R}}
\newcommand{\Z}{\mathbb{Z}}

\newcommand{\ii}{\mathrm{i}}
\newcommand{\e}{\mathrm{e}}

\newcommand{\Jaindent}{\ignorespaces}
\begin{document}

\title{Topological flat bands from nonlinear periodic motion}

\author{Urei Miura}
\email{urei.miura@yukawa.kyoto-u.ac.jp}
\affiliation{Division of Physics and Astronomy, Graduate School of Science,
	Kyoto University, Kyoto 606-8502, Japan}
\affiliation{Center for Gravitational Physics and Quantum Information,
	Yukawa Institute for Theoretical Physics, Kyoto University,
	Kitashirakawa Oiwake-Cho, Sakyo-ku, Kyoto 606-8502, Japan}

\author{Hisao Hayakawa}
\affiliation{Center for Gravitational Physics and Quantum Information,
	Yukawa Institute for Theoretical Physics, Kyoto University,
	Kitashirakawa Oiwake-Cho, Sakyo-ku, Kyoto 606-8502, Japan}


\begin{abstract}
	{\color{black}
		We use the Lax formulation of classical integrable systems to characterize the topology of nonlinear periodic orbits.
		For a Hamiltonian system with one degree of freedom, we express the chiral-symmetric Lax operator along a periodic orbit in a frequency basis. Energy conservation makes both bands of the resulting tight-binding model exactly flat.
		The harmonic oscillator corresponds to a fully dimerized topological Su--Schrieffer--Heeger model, while nonlinearity generates long-range hopping through higher harmonics.
		In the symmetric double well, the winding number changes as intrawell periodic orbits give way to an orbit spanning both wells, and the zero-energy edge mode of the semi-infinite chain disappears.
		We derive the edge-mode profile and show that its localization length diverges logarithmically as the separatrix is approached.
		At the separatrix, the period and the hopping decay length diverge while the band gap remains finite; this singular limit of the periodic motion allows the winding number of the infinite frequency chain to change.
		Canonical quantization of the Lax operator also connects the chiral winding numbers to supersymmetric quantum mechanics: for a bound system on the real line with simple turning points, their sum over all periodic orbits at a common positive energy equals the Witten index.
		We also apply the construction to periodic nonlinear waves through traveling-wave reductions. The finite-period winding number becomes undefined in their solitary-wave limits.
		\par}
\end{abstract}

\maketitle

\section{Introduction}
\label{sec:introduction}
{\color{black}\hypersetup{citecolor=blue}
	\Jaindent Topology characterizes the properties of matter and waves independently of microscopic structural details.
	When the relevant symmetries and bulk gap are preserved, quantized responses and states localized at boundaries remain stable against perturbations. This stability guides both the understanding of material properties and the design of functional systems.
	For linear systems, an established framework computes topological invariants from band eigenstates and classifies systems according to symmetry and dimensionality~\cite{Ryu2010,Kawabata2019}.
	Beyond quantum systems, these ideas have been extended to boundary zero modes in classical spring lattices~\cite{KaneLubensky2014} and geometric phases of vibrational modes~\cite{KariyadoHatsugai2016}.

	\Jaindent Applying topology to nonlinear systems first requires specifying what is to be classified.
	The absence of the superposition principle makes it difficult to organize states using a common set of eigenmodes as in linear systems.
	Moreover, the shape and frequency of a solution depend on its amplitude, and multiple solutions can coexist or undergo bifurcations.
	Linearization describes infinitesimal perturbations.
	To capture nonlinear motion at finite amplitude, one must specify the set of solutions on which an invariant is defined and establish how its value relates to physical responses or edge states. Constructing a classification applicable to a broad range of nonlinear systems remains a challenge.

	\Jaindent Studies have examined topological phases induced by nonlinearity~\cite{Hadad2016} and geometric phases of nonlinear eigenstates~\cite{Tuloup2020}.
	A geometric phase has also been defined for nonsinusoidal periodic waveforms, with quantization enforced by reflection symmetry~\cite{Zhou2022}.
	Nonlinear edge states that preserve chiral symmetry have been studied theoretically~\cite{JezequelDelplace2022} and in acoustic experiments~\cite{Guo2025}.
	A method for defining Chern numbers from nonlinear eigenstates at fixed amplitude has also been proposed~\cite{Sone2024}.
	In mechanical systems, a different approach uses the Poincar\'e--Hopf index to characterize nonlinear zero modes and topologically protected solitons~\cite{Lo2021}.

	\Jaindent Bulk--boundary relations in nonlinear systems can also be studied by treating edge-state profiles as dynamical systems in the spatial direction.
	This approach has shown that the bulk--edge correspondence in a nonlinear Su--Schrieffer--Heeger (SSH) model breaks down when its edge-state profile becomes spatially chaotic~\cite{Sone2025}.
	Onsite nonlinearity can also induce a bifurcation of the spatial dynamical system while preserving the bulk topology, producing a transition in the asymptotic behavior of the edge-state profile~\cite{SoneHatsugai2026}.
	For problems with nonlinear dependence on the eigenfrequency, auxiliary eigenvalues have been introduced to investigate the conditions under which auxiliary bands correspond to physical edge states~\cite{IsobeYoshidaHatsugai2024}.

	\Jaindent Here, we use classical integrable systems to connect nonlinear motion to linear band models.
	Classical integrable systems possess enough conserved quantities to determine their nonlinear motion in principle.
	In particular, a Lax formulation expresses the nonlinear equations of motion as the compatibility condition of an auxiliary linear problem, and the eigenvalues of the Lax operator are conserved under time evolution~\cite{Lax1968}.
	Geometric properties of integrable systems have been studied through the Hannay angle associated with adiabatic changes~\cite{Hannay1985} and the relation between Lax eigenvectors and Maslov indices in the Toda lattice~\cite{FoxmanRobbins2005}.
	For the Toda lattice, both nonlinear edge responses~\cite{EzawaToda2022} and a correspondence between Chern numbers of Lax eigenstates and boundary states~\cite{SatoFukui2023} have been established.
	Building on these studies, we ask how the conserved spectrum and bifurcations of a nonlinear periodic orbit determine the bands and edge states of an associated linear model.

	\Jaindent A frequency representation provides a useful way to convert time dependence into lattice couplings.
	Floquet theory for periodically driven systems uses matrix representations in a harmonic basis~\cite{Shirley1965,Sambe1973}. In multifrequency driving, the topology of the frequency lattice characterizes energy transfer between drives~\cite{Martin2017,Crowley2019}.
	In optical systems, synthetic frequency lattices in which modulation couples different frequency modes have been proposed~\cite{YuanShiFan2016}, and band spectroscopy including long-range hopping has been realized~\cite{Dutt2019}.
	In nonlinear optical frequency conversion, the Fourier coefficients of a periodic pump waveform define the intersublattice couplings of a chiral frequency chain, whose topology is determined by the phase winding of the pump~\cite{Longhi2022}. Dimer chains with long-range couplings have also been observed~\cite{Pellerin2024}.

	\Jaindent Long-range hopping modifies both band topology and dispersion.
	Extending the one-dimensional SSH model~\cite{SSH1979} to long-range couplings allows large winding numbers and multiple edge states while preserving chiral symmetry~\cite{PerezGonzalez2019,LiLongRange2024}.
	Methods for constructing flat bands include approaches based on localized states~\cite{Maimaiti2017}, molecular-orbital representations~\cite{MizoguchiHatsugai2019}, and local unitary transformations that yield models with all bands flat~\cite{Danieli2021}.
	Flat bands and localization in the frequency direction have also been observed in synthetic frequency lattices~\cite{LiFlatBand2022}.
	In a flat band, the group velocity arising from energy dispersion vanishes, allowing the topological properties of the eigenstates to be isolated from dispersion-driven propagation.
	Classical integrable systems with a Lax formulation provide a natural setting for this separation: the Lax eigenvalues remain constant in time while the eigenvectors evolve along a periodic orbit.
	In the frequency representation, the conserved Lax spectrum yields flat bands whose eigenstates retain the topology of this evolution.

	\Jaindent In this paper, we use a two-component chiral-symmetric Lax representation of autonomous Hamiltonian systems with one degree of freedom to construct a linear band model from their periodic orbits.
	The classical dynamics determines the complex off-diagonal element of this operator and its winding over one period.
	Expressing the Lax operator in a frequency basis yields long-range hopping through nonlinearity, while energy conservation fixes the modulus of the off-diagonal element and makes the two bands exactly flat.

	\Jaindent The harmonic oscillator corresponds to the topological phase of a fully dimerized SSH model.
	In the symmetric double well, the bifurcation to a periodic orbit that crosses the barrier changes the winding number and removes the zero-energy edge mode of the semi-infinite chain for a fixed unit cell and boundary termination.
	As the separatrix is approached, the gap remains finite while the period diverges; at the bifurcation, the finite-period winding number is therefore undefined.
	We derive the zero-energy edge-mode profile and show that its localization length diverges logarithmically in harmonic-index units. The hopping decay length diverges as well.

	\Jaindent Canonical quantization of the same Lax operator yields a supersymmetric (SUSY) quantum mechanical system~\cite{Cooper1995}.
	The relation between phase-space winding and chiral zero-mode indices has been formulated more generally through the mode-shell correspondence~\cite{JezequelDelplace2024ModeShell}.
	For bound motion on the real line with simple turning points, we obtain an exact decomposition of the Witten index~\cite{Witten1982Index} into the chiral winding numbers of all periodic orbits at a common positive energy.
	We also apply the construction to periodic nonlinear waves through traveling-wave reductions and examine their solitary-wave limits.
}

\Jaindent This paper is organized as follows. Section~\ref{sec:lax-winding} introduces the Lax formulation, conserved eigenvalues, and winding number.
Section~\ref{sec:orbit-winding} calculates the winding numbers of the harmonic oscillator and the symmetric double well and shows how they differ across a bifurcation of the motion.
Section~\ref{sec:frequency-lattice} constructs the frequency chain and establishes its flatness from energy conservation and the SSH correspondence for the harmonic oscillator.
Section~\ref{sec:nonlinear-transitions} examines long-range hopping, the transition at the separatrix, and edge states.
Section~\ref{sec:susy-witten} establishes the correspondence with the Witten index and its interpretation through semiclassical states and tunneling.
Section~\ref{sec:conclusion} summarizes the results and discusses future directions.

\Jaindent Appendix~\ref{app:turning-point-winding} derives the turning-point formula, and Appendix~\ref{app:frequency-flatness} discusses the hopping cancellation constraints and the hopping range required for flatness.
Appendix~\ref{app:periodic-solutions} presents the periodic solutions and winding numbers of the symmetric double well, power-law oscillators, and the pendulum. Appendix~\ref{app:double-well-edge} examines the Fourier coefficients and zero-energy edge mode of the double well.
Appendix~\ref{app:traveling-wave-reductions} treats applications to nonlinear waves and their solitary-wave limits.

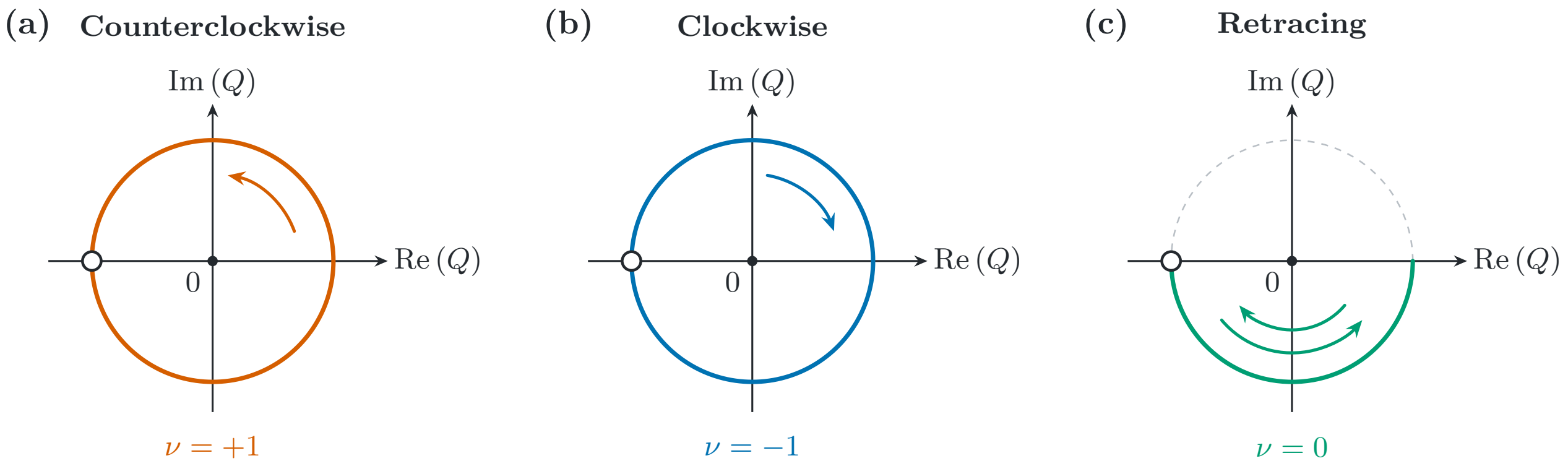
\begin{figure*}[!t]
	\centering
	\resizebox{0.95\textwidth}{!}{%
		\begingroup%
			\tikzset{%
				axis/.style={paper axis,-{Stealth[length=3.8pt,width=3.2pt]}},%
				path/.style={paper curve,draw=trajectory,line width=1.25pt},%
				direction/.style={draw=trajectory,line width=.9pt,line cap=round,%
					-{Stealth[length=5pt,width=4pt]}},%
				startpoint/.style={circle,draw=paperink,fill=white,line width=.8pt,inner sep=1.9pt},%
				origin/.style={circle,fill=paperink,inner sep=1.05pt},%
				panel/.style={font=\PaperPanelFont,anchor=west,inner sep=0pt},%
				every node/.style={font=\PaperLabelFont,text=paperink,inner sep=1.5pt}%
			}%
			\setlength{\fboxsep}{3pt}%
			\setlength{\fboxrule}{0pt}%
			\fbox{%
				\begin{tikzpicture}[x=1cm,y=1cm]
					\path[use as bounding box] (-2.20,-2.05) rectangle (13.6782,2.6252);
					\begin{pgfinterruptboundingbox}
						\begin{scope}
							\colorlet{trajectory}{paperorange}
							\node[panel] at (-2.12,2.38) {(a)};
							\node[font=\PaperLabelFont\bfseries] at (0,2.38) {Counterclockwise};
							\begin{scope}[x=1.22cm,y=1.22cm]
								\draw[axis] (-1.36,0) -- (1.45,0) node[right] {$\operatorname{Re}\qty(Q)$};
								\draw[axis] (0,-1.25) -- (0,1.30) node[above] {$\operatorname{Im}\qty(Q)$};
								\draw[path] plot[domain=180:540,samples=181,variable=\t]
									({cos(\t)},{sin(\t)});
								\draw[direction] ({0.72*cos(20)},{0.72*sin(20)})
									arc[start angle=20,end angle=80,radius=0.72];
								\node[origin] at (0,0) {};
								\node[anchor=north east] at (-0.05,-0.05) {$0$};
								\node[startpoint] at (-1,0) {};
							\end{scope}
							\node[text=trajectory] at (0,-1.88) {$\nu=+1$};
						\end{scope}

						\begin{scope}[xshift=5.45cm]
							\colorlet{trajectory}{paperblue}
							\node[panel] at (-2.12,2.38) {(b)};
							\node[font=\PaperLabelFont\bfseries] at (0,2.38) {Clockwise};
							\begin{scope}[x=1.22cm,y=1.22cm]
								\draw[axis] (-1.36,0) -- (1.45,0) node[right] {$\operatorname{Re}\qty(Q)$};
								\draw[axis] (0,-1.25) -- (0,1.30) node[above] {$\operatorname{Im}\qty(Q)$};
								\draw[path] plot[domain=180:-180,samples=181,variable=\t]
									({cos(\t)},{sin(\t)});
								\draw[direction] ({0.72*cos(80)},{0.72*sin(80)})
									arc[start angle=80,end angle=20,radius=0.72];
								\node[origin] at (0,0) {};
								\node[anchor=north east] at (-0.05,-0.05) {$0$};
								\node[startpoint] at (-1,0) {};
							\end{scope}
							\node[text=trajectory] at (0,-1.88) {$\nu=-1$};
						\end{scope}

						\begin{scope}[xshift=10.90cm]
							\colorlet{trajectory}{paperteal}
							\node[panel] at (-2.12,2.38) {(c)};
							\node[font=\PaperLabelFont\bfseries] at (0,2.38) {Retracing};
							\begin{scope}[x=1.22cm,y=1.22cm]
								\draw[draw=paperguide!60,dash pattern=on 2pt off 2pt,line width=.45pt]
									(0,0) circle[radius=1];
								\draw[axis] (-1.36,0) -- (1.45,0) node[right] {$\operatorname{Re}\qty(Q)$};
								\draw[axis] (0,-1.25) -- (0,1.30) node[above] {$\operatorname{Im}\qty(Q)$};
								\draw[path] (-1,0) arc[start angle=-180,end angle=0,radius=1];
								\draw[direction] ({0.76*cos(220)},{0.76*sin(220)})
									arc[start angle=220,end angle=320,radius=0.76];
								\draw[direction] ({0.57*cos(320)},{0.57*sin(320)})
									arc[start angle=320,end angle=220,radius=0.57];
								\node[origin] at (0,0) {};
								\node[anchor=north east] at (-0.05,-0.05) {$0$};
								\node[startpoint] at (-1,0) {};
							\end{scope}
							\node[text=trajectory] at (0,-1.88) {$\nu=0$};
						\end{scope}
					\end{pgfinterruptboundingbox}
				\end{tikzpicture}%
			}%
		\endgroup%
	}
	\caption{Winding number of $Q\qty(t)$ in the complex plane. A full turn gives (a) $\nu=+1$ for counterclockwise motion and (b) $\nu=-1$ for clockwise motion, while (c) traversing an arc back and forth gives $\nu=0$.
		The arrows inside the circles indicate the direction of time, the open circles mark the initial and final points, and the black dots mark the origin.
		Energy conservation gives $\qty|Q\qty(t)|=\sqrt{2E}$, so all three trajectories lie on circles of the same radius.}
	\label{fig:winding-number-definition}
\end{figure*}

\section{Lax formulation and chiral winding number}
\label{sec:lax-winding}
\Jaindent We apply the Lax formulation to an autonomous Hamiltonian system with one degree of freedom and determine the eigenvalues and winding number along a periodic orbit.

\subsection{Fundamental formulation}
\label{subsec:formulation}
\Jaindent Let $q$ and $p$ denote the canonical coordinate and momentum, respectively. Using a smooth real function $w\qty(q)$, we define an autonomous Hamiltonian with one degree of freedom as
\begin{align}
	H\qty(q,p)\coloneq\frac{1}{2}\qty[p^{2}+w\qty(q)^{2}].
	\label{eq:hamiltonian}
\end{align}
Hamilton's equations are
\begin{align}
	&\dot{q} = p, \notag \\
	&\dot{p} = -w\qty(q)w'\qty(q).
	\label{eq:canonicaleq}
\end{align}
Here, a dot denotes differentiation with respect to time $t$, and a prime denotes differentiation with respect to $q$.
Energy conservation makes this system with one degree of freedom integrable.
We use the following $2\times2$ Lax pair, which is related to a known representation for one-degree-of-freedom systems~\cite{Cardoso2025} by a constant unitary change of basis and a reversal of the commutator convention:
\begin{align}
	& Q\qty(t) \coloneq p - \ii w\qty(q),\label{eq:complex-amplitude} \\
	&\mathsf{L} \coloneq
	\begin{pmatrix}
		0 & Q \\
		Q^{*} & 0
	\end{pmatrix}, \notag \\
	&\mathsf{M} \coloneq -\frac{\ii}{2} w'\qty(q)
	\begin{pmatrix}
		1 & 0 \\
		0 & -1
	\end{pmatrix}, \label{eq:lax-pair} \\
	&\dot{\mathsf{L}} = \comm{\mathsf{M}}{\mathsf{L}}.
	\label{eq:lax-equation}
\end{align}
Hamilton's equations give $\dot Q=-\ii w'\qty(q)Q$, which yields the off-diagonal components of the commutator $\comm{\mathsf M}{\mathsf L}$ and hence the Lax equation \eqref{eq:lax-equation}.
The complex variables $p\pm\ii w\qty(q)$ and their conserved product also appear in classical SUSY mechanics~\cite{JunkerMatthiesen1994}.
The time dependence of $\mathsf{L}$ and $\mathsf{M}$ is determined by the classical solution $q\qty(t),p\qty(t)$.

\Jaindent In terms of the Pauli matrices
\begin{align*}
	&\sigma^{x} \coloneq
	\begin{pmatrix}
		0 & 1 \\
		1 & 0
	\end{pmatrix}, \\
	&\sigma^{y} \coloneq
	\begin{pmatrix}
		0 & -\ii \\
		\ii & 0
	\end{pmatrix}, \\
	&\sigma^{z} \coloneq
	\begin{pmatrix}
		1 & 0 \\
		0 & -1
	\end{pmatrix},
\end{align*}
the same Lax pair can be written as
\begin{align}
	&\mathsf{L} = p\sigma^{x} + w\qty(q)\sigma^{y}, \notag \\
	&\mathsf{M} = -\frac{\ii}{2} w'\qty(q)\sigma^{z}.
	\label{eq:lax-pauli}
\end{align}
Since $\mathsf{L}$ contains only $\sigma^{x}$ and $\sigma^{y}$, it has chiral symmetry:
\begin{align}
	\sigma^{z} \mathsf{L} \sigma^{z} = -\mathsf{L}.
	\label{eq:Lchiralsymmetry}
\end{align}

\Jaindent We denote a classical periodic orbit of period $T$ by $q_{\text{p.s.}}\qty(t),p_{\text{p.s.}}\qty(t)$, where the subscript p.s. stands for periodic solution. Along this orbit,
\begin{align}
	& Q_{\text{p.s.}}\qty(t) = p_{\text{p.s.}}\qty(t) - \ii w\qty(q_{\text{p.s.}}\qty(t)) \notag \\
	&\phantom{Q_{\text{p.s.}}\qty(t)} = Q_{\text{p.s.}}\qty(t+T) \notag \\
	& \mathsf{L}_{\text{p.s.}}\qty(t) =
	\begin{pmatrix}
		0 & Q_{\text{p.s.}}\qty(t) \\
		Q_{\text{p.s.}}\qty(t)^{*} & 0
	\end{pmatrix} \notag \\
	&\phantom{\mathsf{L}_{\text{p.s.}}\qty(t)} = \mathsf{L}_{\text{p.s.}}\qty(t+T).
\end{align}
We define a winding number for this time-periodic $\mathsf{L}$ and express the operator as a lattice model in frequency space in Sec.~\ref{sec:frequency-lattice}.
Hereafter, we omit the subscript p.s. where appropriate.

\subsection{Conserved Lax spectrum}
\label{subsec:flat-band}
\Jaindent The Lax equation preserves the eigenvalues of $\mathsf{L}$. For the present model, let the energy of a periodic orbit be
\begin{align}
	E = H\qty(q_{\text{p.s.}}\qty(t),p_{\text{p.s.}}\qty(t)).
	\label{eq:conserved-energy}
\end{align}
Since $\qty|Q\qty(t)|^{2}=p^{2}+w\qty(q)^{2}=2E$, we obtain
$\mathsf{L}\qty(t)^{2}=2E\mathsf{I}$, where $\mathsf{I}$ is the identity matrix.
The eigenvalues are therefore $\pm\sqrt{2E}$ throughout the period, and $Q$ moves on a circle of radius $\sqrt{2E}$.
Energy conservation fixes this radius, while the waveform of the periodic orbit determines how the phase evolves along the circle.

\begin{figure*}[!t]
	\centering
	\resizebox{0.95\textwidth}{!}{%
		\begingroup%
			\colorlet{outward}{paperblue}%
			\colorlet{returning}{paperorange}%
			\colorlet{allowed}{paperfaint}%
			\tikzset{%
				axis/.style={paper axis,-{Stealth[length=3.8pt,width=3.2pt]}},%
				threshold/.style={draw=paperguide!80,dash pattern=on 3pt off 2pt,line width=.45pt},%
				projection/.style={draw=paperguide!75,densely dotted,line width=.45pt},%
				curve/.style={paper curve,draw=paperink},%
				outward/.style={curve,draw=outward},%
				returning/.style={curve,draw=returning},%
				motion/.style={postaction={decorate},decoration={markings,%
						mark=at position 0.29 with {\arrow{Stealth[length=5.5pt,width=4.5pt]}},%
						mark=at position 0.70 with {\arrow{Stealth[length=5.5pt,width=4.5pt]}}}},%
				openpoint/.style={circle,draw=paperink,fill=white,line width=.8pt,inner sep=1.85pt},%
				filledpoint/.style={circle,draw=paperink,fill=paperink,inner sep=1.6pt},%
				panel/.style={font=\PaperPanelFont,anchor=west,inner sep=0pt},%
				every node/.style={font=\PaperLabelFont,text=paperink,inner sep=1.5pt}%
			}%
			\setlength{\fboxsep}{3pt}%
			\setlength{\fboxrule}{0pt}%
			\fbox{%
				\begin{tikzpicture}[x=1cm,y=1cm]
					\path[use as bounding box] (-2.20,-7.75) rectangle (13.6762,2.7242);
					\begin{pgfinterruptboundingbox}
						\pgfmathsetmacro{\qMinimum}{(-3+sqrt(10.08))/1.8}
						\pgfmathsetmacro{\wMinimum}{1-(1-\qMinimum*\qMinimum)*(1.5+0.3*\qMinimum)}
						\pgfmathsetmacro{\excursion}{-asin(\wMinimum)}

						\node[font=\PaperPanelFont] at (0,2.54) {Signed factor};
						\node[font=\PaperPanelFont] at (5.45,2.54) {Potential};
						\node[font=\PaperPanelFont] at (10.90,2.54) {Lax trajectory};

						\begin{scope}
							\node[panel] at (-2.15,2.18) {(a)\quad $\qty(+,+)$};
							\begin{scope}[x=1.27cm,y=1.02cm]
								\fill[allowed] (-1,-1) rectangle (1,1);
								\draw[threshold] (-1.30,1) -- (1.26,1)
									node[anchor=west] {$\sqrt{2E}$};
								\draw[threshold] (-1.30,-1) -- (1.26,-1)
									node[anchor=west] {$-\sqrt{2E}$};
								\draw[axis] (-1.39,0) -- (1.52,0) node[right] {$q$};
								\draw[axis] (0,-1.14) -- (0,1.60) node[above] {$w\qty(q)$};
								\draw[projection] (-1,-1.17) -- (-1,1);
								\draw[projection] (1,-1.17) -- (1,1);
								\draw[curve] plot[domain=-1.12:1.12,samples=151]
									(\x,{1-(1-\x*\x)*(1.5+0.3*\x)});
								\node[openpoint] at (-1,1) {};
								\node[filledpoint] at (1,1) {};
								\node[below] at (-1,-1.17) {$q_{-}$};
								\node[below] at (1,-1.17) {$q_{+}$};
							\end{scope}
							\draw[outward,-{Stealth[length=5.5pt,width=4.5pt]}] (-1.27,-1.65) -- (1.27,-1.65);
							\draw[returning,-{Stealth[length=5.5pt,width=4.5pt]}] (1.27,-1.89) -- (-1.27,-1.89);
						\end{scope}

						\begin{scope}[xshift=5.45cm]
							\node[panel] at (-2.15,2.18) {(b)};
							\begin{scope}[yshift=-1.19cm,x=1.27cm,y=2.26cm]
								\fill[allowed] (-1,0) rectangle (1,0.5);
								\draw[axis] (-1.39,0) -- (1.52,0) node[right] {$q$};
								\draw[axis] (0,-0.045) -- (0,1.13) node[above] {$w\qty(q)^{2}/2$};
								\draw[threshold] (-1.30,0.5) -- (1.30,0.5) node[right] {$E$};
								\draw[projection] (-1,0) -- (-1,0.5);
								\draw[projection] (1,0) -- (1,0.5);
								\draw[curve] plot[domain=-1.12:1.12,samples=171]
									(\x,{(1-(1-\x*\x)*(1.5+0.3*\x))^2/2});
								\node[openpoint] at (-1,0.5) {};
								\node[filledpoint] at (1,0.5) {};
								\node[below] at (-1,0) {$q_{-}$};
								\node[below] at (1,0) {$q_{+}$};
							\end{scope}
						\end{scope}

						\begin{scope}[xshift=10.90cm]
							\node[panel] at (-2.15,2.18) {(c)};
							\begin{scope}[x=1.19cm,y=1.19cm]
								\draw[draw=paperguide!60,dash pattern=on 2pt off 2pt,line width=.45pt] (0,0) circle[radius=1];
								\draw[axis] (-1.35,0) -- (1.46,0) node[right] {$\operatorname{Re}\qty(Q)$};
								\draw[axis] (0,-1.31) -- (0,1.34) node[above] {$\operatorname{Im}\qty(Q)$};
								\draw[outward] (0,-1) arc[start angle=-90,end angle=\excursion,radius=1];
								\draw[returning] (0,-1) arc[start angle=-90,end angle={-180-\excursion},radius=1];
								\draw[outward,-{Stealth[length=6pt,width=5pt]}]
									({cos(-64)},{sin(-64)}) arc[start angle=-64,end angle=-45,radius=1];
								\draw[outward,-{Stealth[length=6pt,width=5pt]}]
									({cos(24)},{sin(24)}) arc[start angle=24,end angle=7,radius=1];
								\draw[returning,-{Stealth[length=6pt,width=5pt]}]
									({cos(-116)},{sin(-116)}) arc[start angle=-116,end angle=-135,radius=1];
								\draw[returning,-{Stealth[length=6pt,width=5pt]}]
									({cos(-204)},{sin(-204)}) arc[start angle=-204,end angle=-187,radius=1];
								\node[openpoint,inner sep=2.4pt] at (0,-1) {};
								\node[filledpoint,inner sep=1.2pt] at (0,-1) {};
								\node[fill=white,inner sep=3pt] at (0,0.40) {$\nu=0$};
								\node[anchor=north] at (0,-1.38) {$q_{-},q_{+}:\ Q=-\mathrm{i}\sqrt{2E}$};
							\end{scope}
						\end{scope}

						\begin{scope}[yshift=-4.90cm]
							\node[panel] at (-2.15,2.18) {(d)\quad $\qty(-,+)$};
							\begin{scope}[x=1.27cm,y=1.02cm]
								\fill[allowed] (-1,-1) rectangle (1,1);
								\draw[threshold] (-1.30,1) -- (1.26,1)
									node[anchor=west] {$\sqrt{2E}$};
								\draw[threshold] (-1.30,-1) -- (1.26,-1)
									node[anchor=west] {$-\sqrt{2E}$};
								\draw[axis] (-1.39,0) -- (1.52,0) node[right] {$q$};
								\draw[axis] (0,-1.14) -- (0,1.60) node[above] {$w\qty(q)$};
								\draw[projection] (-1,-1.17) -- (-1,1);
								\draw[projection] (1,-1.17) -- (1,1);
								\draw[curve] plot[domain=-1.12:1.30,samples=151]
									(\x,{\x+0.25*(1-\x*\x)});
								\node[openpoint] at (-1,-1) {};
								\node[filledpoint] at (1,1) {};
								\node[below] at (-1,-1.17) {$q_{-}$};
								\node[below] at (1,-1.17) {$q_{+}$};
							\end{scope}
							\draw[outward,-{Stealth[length=5.5pt,width=4.5pt]}] (-1.27,-1.65) -- (1.27,-1.65);
							\draw[returning,-{Stealth[length=5.5pt,width=4.5pt]}] (1.27,-1.89) -- (-1.27,-1.89);
						\end{scope}

						\begin{scope}[xshift=5.45cm,yshift=-4.90cm]
							\node[panel] at (-2.15,2.18) {(e)};
							\begin{scope}[yshift=-1.19cm,x=1.27cm,y=2.26cm]
								\fill[allowed] (-1,0) rectangle (1,0.5);
								\draw[axis] (-1.39,0) -- (1.52,0) node[right] {$q$};
								\draw[axis] (0,-0.045) -- (0,1.13) node[above] {$w\qty(q)^{2}/2$};
								\draw[threshold] (-1.30,0.5) -- (1.30,0.5) node[right] {$E$};
								\draw[projection] (-1,0) -- (-1,0.5);
								\draw[projection] (1,0) -- (1,0.5);
								\draw[curve] plot[domain=-1.30:1.30,samples=171]
									(\x,{(\x+0.25*(1-\x*\x))^2/2});
								\node[openpoint] at (-1,0.5) {};
								\node[filledpoint] at (1,0.5) {};
								\node[below] at (-1,0) {$q_{-}$};
								\node[below] at (1,0) {$q_{+}$};
							\end{scope}
						\end{scope}

						\begin{scope}[xshift=10.90cm,yshift=-4.90cm]
							\node[panel] at (-2.15,2.18) {(f)};
							\begin{scope}[x=1.19cm,y=1.19cm]
								\draw[axis] (-1.35,0) -- (1.46,0) node[right] {$\operatorname{Re}\qty(Q)$};
								\draw[axis] (0,-1.31) -- (0,1.34) node[above] {$\operatorname{Im}\qty(Q)$};
								\draw[outward,motion] plot[domain=90:-90,samples=101,variable=\t]
									({cos(\t)},{sin(\t)});
								\draw[returning,motion] plot[domain=-90:-270,samples=101,variable=\t]
									({cos(\t)},{sin(\t)});
								\node[openpoint] at (0,1) {};
								\node[filledpoint] at (0,-1) {};
								\node[anchor=east] at (-0.12,1.05) {$q_{-}$};
								\node[anchor=east] at (-0.12,-1.05) {$q_{+}$};
								\node[fill=white,inner sep=3pt] at (0,0.35) {$\nu=-1$};
							\end{scope}
						\end{scope}

						\draw[outward,-{Stealth[length=6pt,width=5pt]}] (-1.95,-7.45) -- (-1.33,-7.45);
						\node[anchor=west] at (-1.19,-7.45) {$p>0:\ q_{-}\to q_{+}$};
						\draw[returning,-{Stealth[length=6pt,width=5pt]}] (4.25,-7.45) -- (3.63,-7.45);
						\node[anchor=west] at (4.39,-7.45) {$p<0:\ q_{+}\to q_{-}$};
						\node[anchor=west] at (9.13,-7.45) {$\qty|Q|=\sqrt{2E}$};
					\end{pgfinterruptboundingbox}
				\end{tikzpicture}%
			}%
		\endgroup%
	}
	\caption{Signs of $w$ at the turning points and the winding number. From left to right, the panels show $w\qty(q)$, the potential $w\qty(q)^{2}/2$, and the trajectory of $Q$.
		Intersections with the horizontal dashed lines give the turning points $q_{-},q_{+}$, and the shaded regions indicate the intervals of $q$ traversed by the periodic orbits. Blue and orange denote the forward and return paths, respectively, and the arrows indicate the direction of time.
		In the top row, the endpoint signs are $\qty(+,+)$, giving $\nu=0$ even though $w$ changes sign within the interval.
		In the bottom row, the signs are $\qty(-,+)$, and $Q$ makes one clockwise turn, giving $\nu=-1$.}
	\label{fig:turning-point-winding}
\end{figure*}
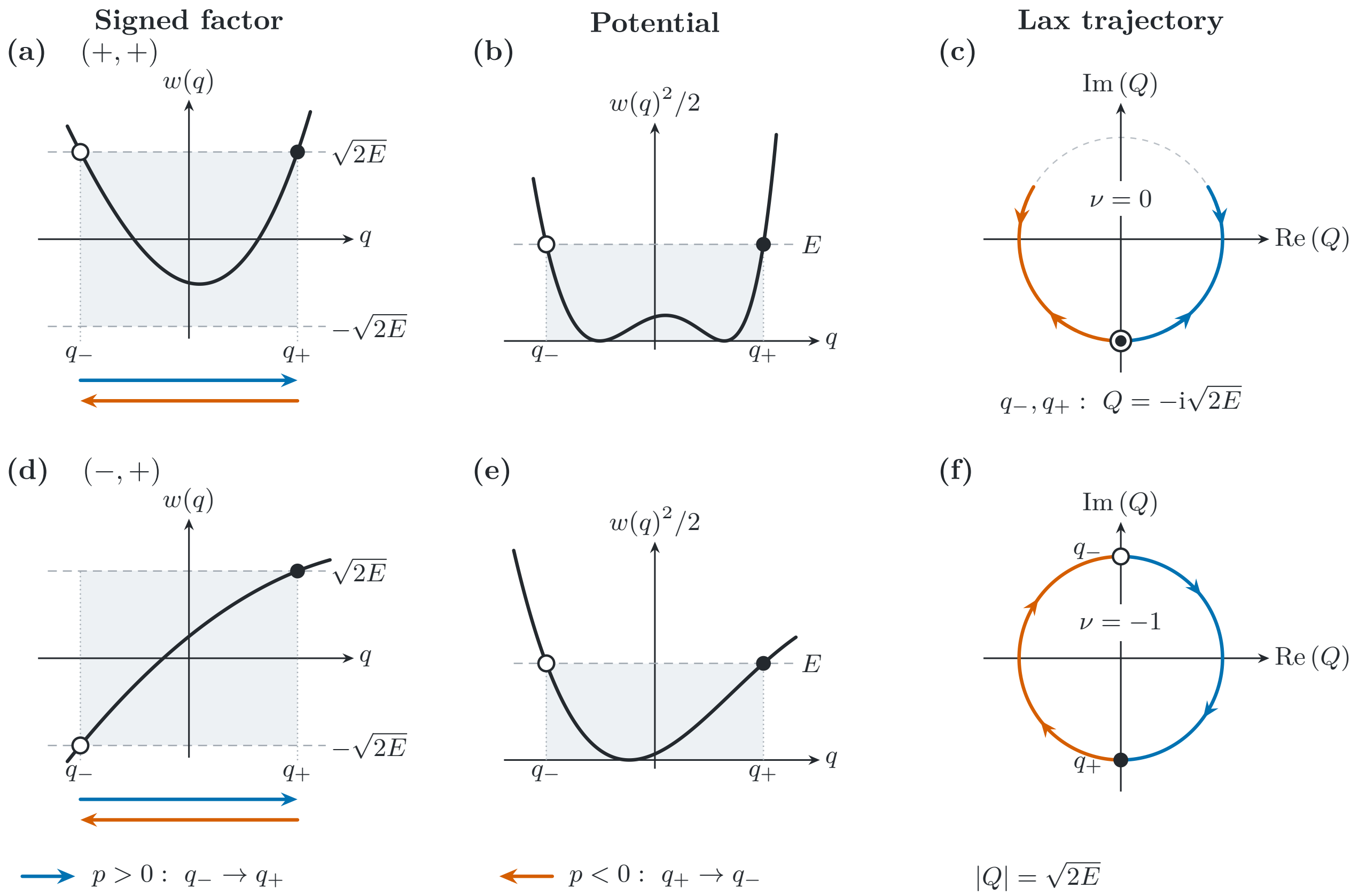

\subsection{Winding number protected by the chiral symmetry}
\label{subsec:chiral-winding}
\Jaindent We fix this Lax representation and a time-independent sublattice basis and examine how the winding number varies among periodic orbits.
For $E>0$, $Q$ traces a closed curve over one period without passing through the origin (Fig.~\ref{fig:winding-number-definition}).
The winding number of this curve is an integer that characterizes a family of chiral-symmetric $2\times2$ matrices~\cite{Ryu2010}.
We define it as
\begin{align}
	&\nu \coloneq \frac{1}{2\pi\ii}\int_{0}^{T}\dd{t}\,
	\dv{t}\ln \qty[Q\qty(t)] \in \Z.
	\label{eq:winding-number}
\end{align}
Here, the phase of $Q$ entering the logarithm is chosen continuously along the periodic orbit.
The integer $\nu$ counts the number of turns around the origin, with counterclockwise turns taken as positive.
Reversing the direction of a full turn reverses the sign, whereas traversing an arc back and forth gives zero.
This integer is preserved under continuous deformations of the closed curve that avoid the origin.

\Jaindent The equation of motion for $Q$ gives the phase velocity $-w'\qty(q\qty(t))$.
The winding number can therefore also be written as
\begin{align}
	&\nu = -\frac{1}{2\pi}\int_{0}^{T}\dd{t}\,
	w'\qty(q\qty(t)).
	\label{eq:winding-potential}
\end{align}

\Jaindent In particular, for a nonstationary periodic orbit that oscillates between two adjacent simple turning points $q_{-}<q_{+}$ on the real line, the winding number over the fundamental period $T$ is
\begin{align}
	&\nu = \frac{\operatorname{sgn}\qty[w\qty(q_{-})]-\operatorname{sgn}\qty[w\qty(q_{+})]}{2}.
	\label{eq:winding-turning-points}
\end{align}
Here, $\operatorname{sgn}$ denotes the sign function.
This relation, which expresses the phase accumulated over one period in terms of the turning points, also appears in discussions of phase in classical SUSY mechanics~\cite{InomataJunker1994,HeumannManton2000}.
A derivation using our conventions is given in Appendix~\ref{app:turning-point-winding}.

\Jaindent The signs of $w$ at the two turning points determine $\nu\in\qty{-1,0,+1}$ (Fig.~\ref{fig:turning-point-winding}).
The transformation $w\mapsto-w$ leaves the Hamiltonian unchanged and complex conjugates $Q$, reversing its winding direction.
Thus, $\nu$ characterizes a periodic orbit within a Lax representation for which the signed function $w$, sublattice basis, and direction of time are fixed.

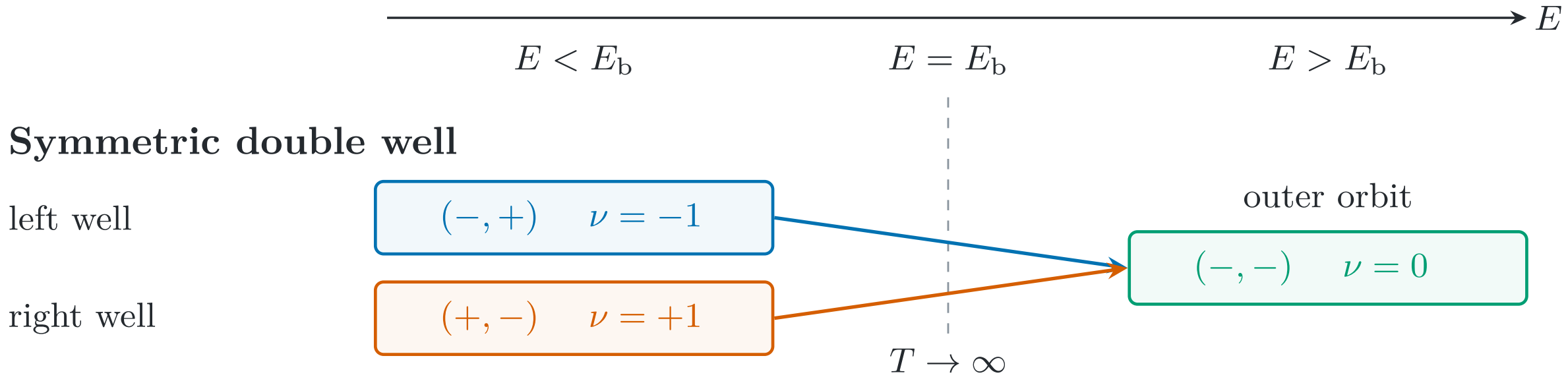
\begin{figure*}[!t]
	\centering
	\resizebox{0.9\textwidth}{!}{%
		\begingroup%
			\colorlet{minus}{paperblue}%
			\colorlet{plus}{paperorange}%
			\colorlet{zero}{paperteal}%
			\newcommand{\branchlabel}[2]{%
				\makebox[1.2cm][c]{$\qty(#1)$}\hspace{.25cm}%
				\makebox[1.2cm][l]{$\nu=#2$}}%
			\tikzset{%
				branch/.style={rounded corners=2pt,minimum width=3.4cm,minimum height=.62cm,%
					inner sep=4pt,align=center,line width=.75pt},%
				negative/.style={branch,draw=minus,fill=minus!5,text=minus},%
				positive/.style={branch,draw=plus,fill=plus!5,text=plus},%
				trivial/.style={branch,draw=zero,fill=zero!5,text=zero},%
				flow/.style={-{Stealth[length=5pt,width=4pt]},line width=.85pt,draw=paperink},%
				every node/.style={font=\PaperLabelFont,text=paperink,inner sep=1.5pt}%
			}%
			\setlength{\fboxsep}{3pt}%
			\setlength{\fboxrule}{0pt}%
			\fbox{%
				\begin{tikzpicture}[x=1cm,y=1cm]
					\path[use as bounding box] (-.05,.30) rectangle (13.4982,3.6692);
					\begin{pgfinterruptboundingbox}
						\node[anchor=west,font=\PaperPanelFont] at (0,2.40) {Symmetric double well};
						\node at (4.9,3.12) {$E<E_{\mathrm{b}}$};
						\node at (8.10,3.12) {$E=E_{\mathrm{b}}$};
						\node at (11.35,3.12) {$E>E_{\mathrm{b}}$};
						\draw[paper axis,-{Stealth[length=4pt,width=3.5pt]}] (3.3,3.48) -- (13.05,3.48) node[right] {$E$};
						\draw[paper guide,dash pattern=on 2.5pt off 2.5pt] (8.10,2.80) -- (8.10,.55);
						\node[fill=white,inner sep=3pt] at (8.10,.55) {$T\to\infty$};

						\node[anchor=west] at (0,1.77) {left well};
						\node[anchor=west] at (0,.91) {right well};
						\node[negative] (sl) at (4.9,1.77) {\branchlabel{-,+}{-1}};
						\node[positive] (sr) at (4.9,.91) {\branchlabel{+,-}{+1}};
						\node[trivial] (so) at (11.35,1.34) {\branchlabel{-,-}{0}};
						\node[anchor=south] at (11.35,1.80) {outer orbit};
						\draw[flow,draw=minus] (sl.east) -- (so.west);
						\draw[flow,draw=plus] (sr.east) -- (so.west);

					\end{pgfinterruptboundingbox}
				\end{tikzpicture}%
			}%
		\endgroup%
	}
	\caption{Bifurcation of periodic orbits and winding numbers in the symmetric double well. The signs of $w$ at the two turning points are shown in parentheses.
		As the energy increases, intrawell periodic orbits give way to an orbit spanning both wells. The period diverges on the dashed line.}
	\label{fig:double-well-branches}
\end{figure*}

\section{Winding numbers of the harmonic oscillator and the double well}
\label{sec:orbit-winding}
\Jaindent Taking the harmonic oscillator as a reference, we examine how anharmonicity and bifurcations of the motion affect the winding number in the symmetric double well.

\subsection{Harmonic oscillator}
\label{subsec:harmonic-oscillator}
\Jaindent We first consider the harmonic oscillator
\begin{align}
	H\qty(q,p)\coloneq \frac{1}{2}\qty[p^{2}+q^{2}]
\end{align}
and choose $w\qty(q)=q$.
For energy $E>0$, taking the right turning point at $t=0$ gives
\begin{align}
	& q\qty(t)= \sqrt{2E}\cos\qty(t) \notag \\
	& p\qty(t)= -\sqrt{2E}\sin\qty(t).
\end{align}
Thus,
\begin{align}
	&Q\qty(t) = p\qty(t)-\ii q\qty(t)
	= -\ii\sqrt{2E}\e^{-\ii t}.
	\label{eq:harmonic-Q}
\end{align}
Since $Q$ makes one clockwise turn at a constant speed, the winding number \eqref{eq:winding-number} over the fundamental period $T=2\pi$ is
\begin{align}
	&\nu = -1.
	\label{eq:harmonic-winding}
\end{align}
This value is preserved for all $E>0$.
For the stationary solution at $E=0$, $Q=0$ and the winding number is undefined.

\subsection{Symmetric double well}
\label{subsec:symmetric-double-well}
\Jaindent Next, we consider the symmetric double well defined by
\begin{align}
	w\qty(q)\coloneq1-q^{2}.
	\label{eq:symmetric-w}
\end{align}
The potential $\qty(1-q^{2})^{2}/2$ has minima at $q=\pm1$, and the barrier at $q=0$ separating the two wells has height $E_{\mathrm{b}}=1/2$.
The particle oscillates within one well for $0<E<1/2$ and across both wells for $E>1/2$.
These motions are described by the Jacobi elliptic dn and cn solutions, respectively.
The periodic orbits are shown in Fig.~\ref{fig:double-well-branches}, and explicit solutions and periods are given in Appendix~\ref{app:periodic-solutions}.

\Jaindent In the left well, $w=-\sqrt{2E}$ at the outer left turning point and $w=+\sqrt{2E}$ at the inner right turning point; this order is reversed in the right well.
For oscillations across both wells, $w=-\sqrt{2E}$ at both turning points.
The turning-point formula \eqref{eq:winding-turning-points} therefore gives
\begin{align}
	\nu=\begin{cases}
		-1, & 0<E<1/2\quad\text{(left well)},\\
		+1, & 0<E<1/2\quad\text{(right well)},\\
		0, & E>1/2
	\end{cases}.
	\label{eq:symmetric-winding}
\end{align}

\Jaindent In the small-amplitude limit $E\to+0$, the periodic orbits in each well approach harmonic oscillations.
Since $w'\qty(-1)=2$ and $w'\qty(+1)=-2$, $Q$ makes one clockwise turn in the left well and one counterclockwise turn in the right well.
As the amplitude increases, $q$ retains its sign for the intrawell dn solution, so the direction of rotation given by $\dot Q=2\ii qQ$ is preserved.
For the cn solution spanning both wells, $q$ changes sign, and the rotations in opposite directions cancel over one period.
In phase space, two periodic orbits related by the reflection $\qty(q,p)\mapsto\qty(-q,-p)$ merge into a single periodic orbit encircling both wells.

\Jaindent At $E=1/2$, the motion follows a separatrix orbit. The period $T$ diverges, and the finite-period winding number is undefined.
The periodic orbits on either side have different winding numbers, as given in Eq.~\eqref{eq:symmetric-winding}.

\Jaindent The winding numbers of all periodic orbits at the same energy sum to zero both below and above the barrier.
The correspondence between this sum and the Witten index of the supersymmetric quantum mechanical system is established in Sec.~\ref{sec:susy-witten}.
Table~\ref{tab:periodic-orbits} compares the turning-point signs and winding numbers of the double well with the power-law and pendulum examples discussed in Appendix~\ref{app:periodic-solutions}.

\begin{table*}[!t]
	\centering
	\caption{Representative periodic orbits and their winding numbers. Orange shading indicates the interval of $q$ traversed by the periodic orbit, and black dots mark the turning points. Dashed lines denote $E$ in the potential plots and $\pm\sqrt{2E}$ in the plots of $w\qty(q)$.
		The signs are listed in the order of the left and right turning points. The power-law examples use odd $n=3$ and even $n=2$.
		The barrier height of the symmetric double well is $E_{\mathrm{b}}=1/2$.
		For pendulum rotation, one period comprises two rotations, over which $Q$ closes.}
	\label{tab:periodic-orbits}
	\begingroup
		\small
		\setlength{\tabcolsep}{4pt}
		\renewcommand{\arraystretch}{1.12}
		\begin{ruledtabular}
\begin{tabular}{p{30mm}p{33mm}cccc}
				Model & \shortstack[l]{Periodic orbit\\Energy} & \shortstack{Potential\\$w\qty(q)^{2}/2$} & $w\qty(q)$ & \shortstack{Signs of\\$w\qty(q_{-}),w\qty(q_{+})$} & $\nu$ \\
\end{tabular}
		\end{ruledtabular}
	\endgroup
\end{table*}

\section{Tight-binding models in frequency space}
\label{sec:frequency-lattice}
\begin{figure*}[!t]
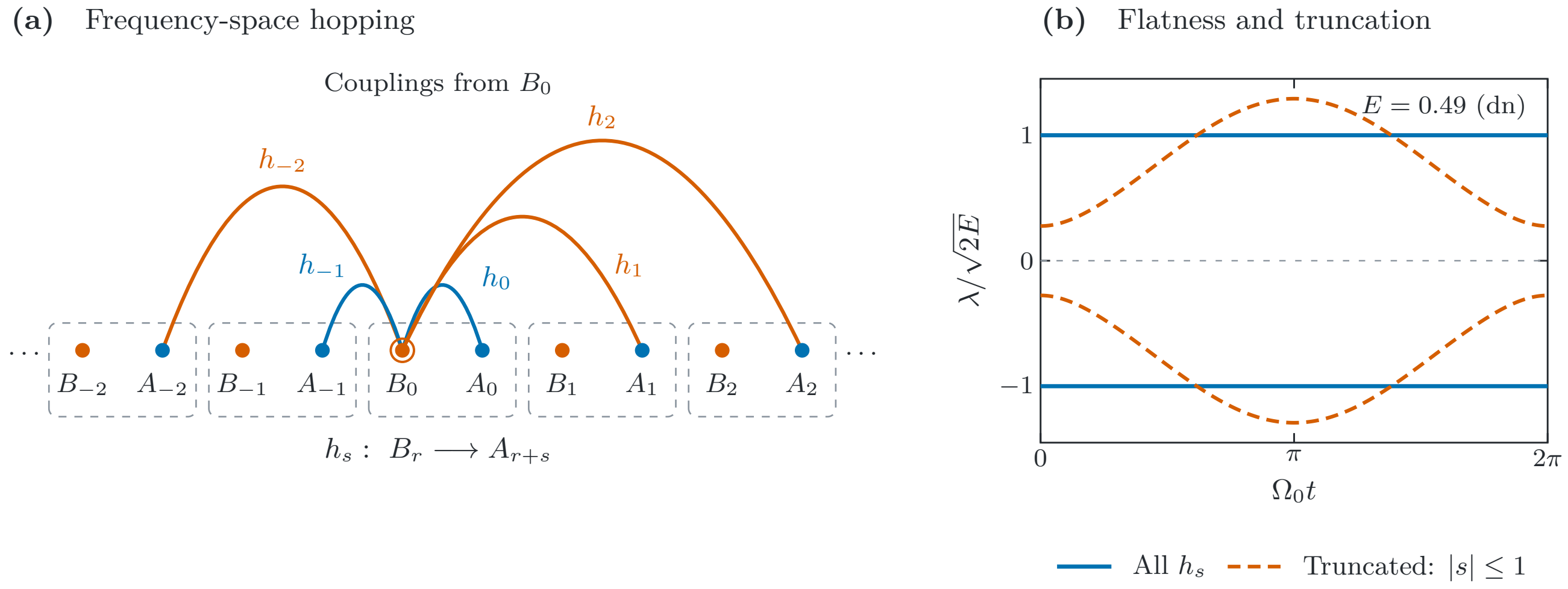

	\centering
	\resizebox{175mm}{!}{%
		\begingroup%
			\pgfplotsset{paper plot/.style={
					scale only axis,axis line style={paper axis},tick align=inside,tick pos=both,
					tick style={draw=paperink,line width=.55pt},major tick length=3pt,
					tick label style={font=\fontsize{8}{9.6}\selectfont,inner sep=2pt},
					label style={font=\fontsize{9}{10.8}\selectfont,inner sep=2pt},
					title style={font=\PaperLabelFont},scaled ticks=false,
					legend style={font=\fontsize{8.2}{9.84}\selectfont,draw=none,fill=none},
					every axis plot/.append style={line width=1.15pt,no marks}}}%

		\endgroup%
	}
	\caption{Couplings in the frequency chain and flatness imposed by energy conservation.
		(a) Representative couplings from $B_{0}$ to $A_{s}$. The reverse hopping amplitude is $h_{s}^{*}$, and the dashed box marks one unit cell.
		Blue denotes $h_{-1},h_{0}$, and orange denotes the remaining long-range couplings.
		(b) Eigenvalues in the time representation for the dn solution in the left well of the symmetric double well at $E=0.49$.
		Solid lines include all hopping terms, while dashed lines show the result of truncation to $\qty|s|\le1$.
		Retaining all components gives flat bands at $\lambda_{\pm}/\sqrt{2E}=\pm1$.}
	\label{fig:frequency-chain-flatness}
\end{figure*}

\Jaindent In a Floquet quasienergy problem, the harmonic representation of periodic coefficients is supplemented by a diagonal term proportional to the harmonic index, arising from the time derivative~\cite{Shirley1965,Sambe1973}.
Here, we express the time-periodic Lax operator $\mathsf{L}\qty(t)$ itself in a frequency basis to construct a tight-binding model in which each harmonic defines a site and the Fourier coefficients of $Q$ give the hopping amplitudes, as illustrated in Fig.~\ref{fig:frequency-chain-flatness} (a).

\subsection{From the Lax operator to a frequency lattice}
\label{subsec:frequency-flat-bands}

\Jaindent For a periodic orbit with $E>0$, let $T$ be the period over which $Q\qty(t)$ and $\mathsf L\qty(t)$ close, and let $\Omega_{0}=2\pi/T$ be the fundamental angular frequency.
Denoting the two components of the Lax operator by $A,B$, we introduce the operator $\hat H$ obtained from $\mathsf L\qty(t)$ in the time basis:
\begin{align}
	&\hat H\coloneq\frac{1}{T}\int_{0}^{T}\dd{t}\,
	\qty(\ket{t,A},\ket{t,B})\,\mathsf L\qty(t)\,
	\qty(\begin{matrix}\bra{t,A}\\\bra{t,B}\end{matrix}).
	\label{eq:frequency-time-hamiltonian}
\end{align}
Choosing a classical periodic orbit fixes the coefficients of $\hat H$.
We expand the time basis in an orthonormal frequency basis as
\begin{align}
	&\ket{t,A}=\sum_{r\in\Z}\e^{-\ii r\Omega_{0}t}\ket{r\Omega_{0},A},\notag\\
	&\ket{t,B}=\sum_{r\in\Z}\e^{-\ii r\Omega_{0}t}\ket{r\Omega_{0},B}.
	\label{eq:frequency-time-basis}
\end{align}
The two components $A,B$ with the same harmonic index $r$ form a unit cell.
The intersublattice matrix elements of Eq.~\eqref{eq:frequency-time-hamiltonian} in this basis are
\begin{align}
	\bra{r\Omega_{0},A}\hat H\ket{r'\Omega_{0},B}
	&=\frac{1}{T}\int_{0}^{T}\dd{t}\,
	\e^{-\ii\qty(r-r')\Omega_{0}t}Q\qty(t).
	\label{eq:frequency-matrix-element}
\end{align}
Defining the Fourier coefficients by
\begin{align}
	&h_{s}\coloneq\frac{1}{T}\int_{0}^{T}\dd{t}\,
	\e^{-\ii s\Omega_{0}t}Q\qty(t),\notag\\
	&Q\qty(t)=\sum_{s\in\Z}h_{s}\e^{\ii s\Omega_{0}t},
	\label{eq:frequency-fourier}
\end{align}
we can write the model in frequency space as
\begin{align}
	&\hat H=\sum_{r,s\in\Z}\qty[
		h_{s}\ket{\qty(r+s)\Omega_{0},A}\bra{r\Omega_{0},B}
		+\mathrm{H.c.}].
	\label{eq:frequency-hamiltonian}
\end{align}
Here, $\mathrm{H.c.}$ denotes the Hermitian conjugate.
Writing the sites as $A_{r},B_{r}$, we see that $h_{s}$ couples $B_{r}$ to $A_{r+s}$, with $h_{0}$ giving the intracell coupling.
Higher harmonics generate hopping to more distant unit cells.
Since every coupling connects different sublattices, the model retains chiral symmetry, with the chiral operator acting as $+1,-1$ on $A,B$, respectively.
The same relation between a periodic waveform, its Fourier coefficients, and a chiral frequency chain appears in optical frequency conversion~\cite{Longhi2022}.
Here, the waveform $Q\qty(t)$ is determined by a classical periodic orbit, and its modulus is fixed by energy conservation.

\subsection{Energy conservation and flatness}
\Jaindent Let $\lambda_{\pm}\qty(t)$ denote the two eigenvalues of $\mathsf L\qty(t)$.
Energy conservation, $\abs{Q}^{2}=2E$, gives
\begin{align}
	&\mathsf L\qty(t)^{2}=2E\mathsf{I},\qquad
	\lambda_{\pm}\qty(t)=\pm\sqrt{2E},
	\label{eq:frequency-flat-spectrum}
\end{align}
so the two bands are flat.
Thus, the conserved Lax spectrum directly supplies a flat representative of a chiral band model, whose topology is conventionally studied using spectral flattening~\cite{Ryu2010}.

\Jaindent In terms of the hopping amplitudes, this flatness is expressed as
\begin{align}
	&\sum_{r\in\Z}h_{r+s}h_{r}^{*}=2E\delta_{s,0}.
	\label{eq:frequency-flatness-constraint}
\end{align}
For $s=0$, the sum of the squared moduli of the coupling amplitudes is fixed. For $s\ne0$, two-step hopping between distinct unit cells cancels when summed over all paths.
Energy conservation relates all the couplings generated by higher harmonics through this cancellation constraint.
Consequently, $\hat H^{2}=2E$ for the chain without boundaries, and its two bands are exactly flat even in the presence of long-range hopping.
The derivation is given in Appendix~\ref{app:frequency-flatness}.

\Jaindent Hereafter, we use long-range hopping to refer to hopping beyond the nearest neighbors of the SSH chain.
The corresponding amplitudes obtained from the elliptic-function solutions decay exponentially with distance.

\Jaindent Truncating the hopping to $\qty|s|\le1$ leaves a residual in the cancellation of two-step hopping processes and produces band dispersion.
Figure~\ref{fig:frequency-chain-flatness} (b) compares the eigenvalues obtained from the dn solution in the left well with all hopping terms retained and with the hopping truncated.

\subsection{Fully dimerized topological SSH limit}
\label{subsec:harmonic-ssh}
\Jaindent For the harmonic oscillator, taking the right turning point as the time origin, as in Sec.~\ref{subsec:harmonic-oscillator}, gives
\begin{align}
	&Q\qty(t)=-\ii\sqrt{2E}\e^{-\ii\Omega_{0}t},\qquad\Omega_{0}=1,\notag\\
	&h_{s}=-\ii\sqrt{2E}\delta_{s,-1}.
	\label{eq:harmonic-fourier-coefficient}
\end{align}
Thus, $h_{-1}$ is the only nonzero coefficient.
Substituting into Eq.~\eqref{eq:frequency-hamiltonian}, we obtain
\begin{align}
	&\hat H=\sum_{r\in\Z}\qty[
		-\ii\sqrt{2E}\ket{r\Omega_{0},A}\bra{\qty(r+1)\Omega_{0},B}
		+\mathrm{H.c.}].
	\label{eq:harmonic-frequency-hamiltonian}
\end{align}
The basis change $\ket{r\Omega_{0},B}\mapsto\ii\ket{r\Omega_{0},B}$ on sublattice $B$ makes the coupling amplitude real and positive, with value $\sqrt{2E}$.
This is the SSH model~\cite{SSH1979} with
\begin{align}
	&h_{0}=0,\qquad \qty|h_{-1}|=\sqrt{2E}.
	\label{eq:harmonic-ssh-couplings}
\end{align}
It is the fully dimerized limit with only intercell couplings and corresponds to the topological phase with $\nu=-1$ for the present unit cell and Fourier convention.

\Jaindent Keeping the same unit cell and retaining $r\ge0$ yields a semi-infinite chain consisting of independent dimers formed by $A_{r}$ and $B_{r+1}$ and an uncoupled edge site $B_{0}$ (Fig.~\ref{fig:harmonic-ssh}).
Each dimer has eigenvalues $\pm\sqrt{2E}$, while $\ket{0\Omega_{0},B}$ is a zero mode localized on a single edge site, corresponding to the winding number~\cite{RyuHatsugai2002}.
We use this unit cell and termination to examine edge states in the next section.

\begin{figure}[!htb]
	\centering
	\resizebox{\columnwidth}{!}{%
		\begingroup%
			\colorlet{acolor}{paperblue}%
			\colorlet{bcolor}{paperorange}%
			\colorlet{cellgray}{paperguide!80}%
			\tikzset{%
				every node/.style={font=\PaperLabelFont,text=paperink,inner sep=1pt},%
				asite/.style={circle,draw=acolor,fill=acolor,line width=0.65pt,minimum size=5pt,inner sep=0pt},%
				bsite/.style={circle,draw=bcolor,fill=bcolor,line width=0.65pt,minimum size=5pt,inner sep=0pt},%
				bond/.style={draw=paperink,line width=1.35pt,line cap=round},%
				cell/.style={draw=cellgray,dash pattern=on 2pt off 2pt,line width=0.45pt,rounded corners=2pt}%
			}%
			\setlength{\fboxsep}{3pt}%
			\setlength{\fboxrule}{0pt}%
			\fbox{%
				\begin{tikzpicture}[x=1.12cm,y=1cm]
					\draw[bond] (0.76,0) -- (1.70,0);
					\draw[bond] (2.46,0) -- (3.40,0);
					\draw[bond] (4.16,0) -- (5.10,0);
					\draw[bond] (5.86,0) -- (6.50,0);
					\foreach \x/\n in {0/0,1.70/1,3.40/2,5.10/3}{
						\draw[cell] (\x-0.40,-0.43) rectangle (\x+1.16,0.47);
						\node[bsite] at (\x,0) {};
						\node[asite] at (\x+0.76,0) {};
						\node[anchor=north] at (\x,-0.12) {$B_{\n}$};
						\node[anchor=south] at (\x+0.76,0.12) {$A_{\n}$};
					}
					\node at (6.78,0) {$\cdots$};
					\node[anchor=south] at (3.30,0.61) {$h_{0}=0,\qquad \qty|h_{-1}|=\sqrt{2E}$};
					\draw[draw=bcolor,line width=0.8pt] (0,0) circle[radius=0.13];
					\node[anchor=north] at (0,-0.60) {$\lambda=0$};
					\node[anchor=north] at (3.30,-0.60) {$\nu=-1,\qquad\lambda_{\pm}=\pm\sqrt{2E}$};
					\pgfresetboundingbox
					\path[use as bounding box] (-12.97157pt,-28.6812pt) rectangle (223.61256pt,29.0545pt);
				\end{tikzpicture}%
			}%
		\endgroup%
	}
	\caption{Semi-infinite chain of the fully dimerized SSH model corresponding to the harmonic oscillator.
		The sites $A_{r},B_{r}$ denote $\ket{r\Omega_{0},A},\ket{r\Omega_{0},B}$, respectively, and the dashed box encloses one unit cell.
		Only $A_{r}$ and $B_{r+1}$ are coupled, and a zero mode is localized at the uncoupled left edge $B_{0}$. The chain extends infinitely to the right.}
	\label{fig:harmonic-ssh}
\end{figure}
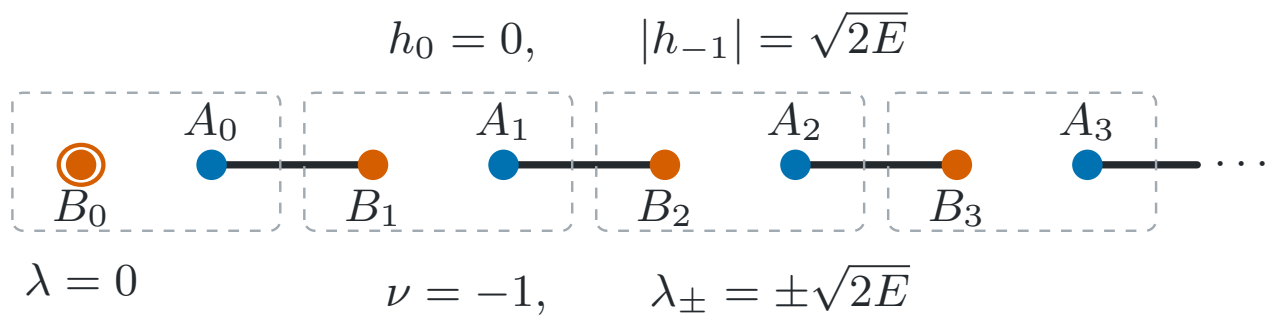

\begin{figure*}[p]
	\centering
	\resizebox{175mm}{!}{%
		\begingroup%
			\pgfplotsset{paper plot/.style={
					scale only axis,clip=false,axis line style={paper axis},tick align=inside,tick pos=both,
					tick style={draw=paperink,line width=.55pt},major tick length=3pt,
					tick label style={font=\fontsize{8}{9.6}\selectfont,inner sep=2pt},
					label style={font=\fontsize{9}{10.8}\selectfont,inner sep=2pt},
					title style={font=\PaperLabelFont},scaled ticks=false,
					legend style={font=\fontsize{8.2}{9.84}\selectfont,draw=none,fill=none},
					every axis plot/.append style={line width=1.15pt}}}%
			\begin{tikzpicture}[x=1mm,y=1mm,every node/.style={font=\PaperLabelFont,text=paperink,inner sep=0pt}]
				\path[use as bounding box] (0,0) rectangle (175,192);
				\begin{pgfinterruptboundingbox}
					\draw[paperblue,line width=1.15pt] (58,187.5)--(65,187.5);\fill[paperblue] (61.5,187.5) circle[radius=1.75pt];
					\node[anchor=west,font=\fontsize{8.2}{9.84}\selectfont] at (67,187.5) {Nearest-neighbor: $s=-1,0$};
					\draw[paperorange,line width=1.15pt] (108,187.5)--(115,187.5);\fill[paperorange] (111.5,187.5) circle[radius=1.75pt];
					\node[anchor=west,font=\fontsize{8.2}{9.84}\selectfont] at (117,187.5) {Long-range};
					\begin{axis}[paper plot,at={(14.875000000mm,147.484285714mm)},anchor=south west,width=46.59763314mm,height=25.135714286mm,
							xmin=-6.6,xmax=6.6,ymin=-.025,ymax=1.075,xtick={-6,-3,0,3,6},minor xtick={-5,-4,-2,-1,1,2,4,5},
							ytick={0,.5,1},yticklabels={$0.0$,$0.5$,$1.0$},xlabel={$s$},ylabel={$\qty|h_{s}|/\sqrt{2E}$},title={Harmonic oscillator}]
						\addplot[paperink,line width=.45pt,no marks] coordinates {(-6.6,0) (6.6,0)};
						\addplot[ycomb,paperblue,line width=1.35pt,mark=*,mark size=1.85pt,mark options={solid,fill=paperblue}] coordinates {
							(-1,1)
						};
						\node[anchor=south west,font=\PaperPanelFont] at (axis description cs:0,1.20) {(a)};
						\node[anchor=north east,font=\fontsize{8}{9.6}\selectfont] at (axis description cs:.96,.83) {$h_{-1}$ only};
					\end{axis}
					\begin{axis}[paper plot,at={(70.326183430mm,147.484285714mm)},anchor=south west,width=46.59763314mm,height=25.135714286mm,
							xmin=-6.6,xmax=6.6,ymin=-.025,ymax=1.075,xtick={-6,-3,0,3,6},minor xtick={-5,-4,-2,-1,1,2,4,5},
							ytick={0,.5,1},yticklabels={$0.0$,$0.5$,$1.0$},xlabel={$s$},yticklabels={},title={$E=0.10$ (dn)}]
						\addplot[paperink,line width=.45pt,no marks] coordinates {(-6.6,0) (6.6,0)};
						\addplot[ycomb,paperblue,line width=1.35pt,mark=*,mark size=1.85pt,mark options={solid,fill=paperblue}] coordinates {
							(-1,0.9858049655591926)
							(0,0.1188799775834082)
						};
						\addplot[ycomb,paperorange,line width=1.35pt,mark=*,mark size=1.85pt,mark options={solid,fill=paperorange}] coordinates {
							(-6,4.547612386674212e-06)
							(-5,6.330424613745122e-05)
							(-4,0.0008459671028969523)
							(-3,0.010598523936733)
							(-2,0.1180279251693664)
							(1,0.003532886686250118)
							(2,1.515870795863386e-06)
							(3,4.878221551945875e-10)
							(4,1.395643746817013e-13)
							(5,8.455252803384993e-17)
							(6,2.823034942775959e-17)
						};
						\node[anchor=south west,font=\PaperPanelFont] at (axis description cs:0,1.20) {(b)};
					\end{axis}
					\begin{axis}[paper plot,at={(125.777366860mm,147.484285714mm)},anchor=south west,width=46.59763314mm,height=25.135714286mm,
							xmin=-6.6,xmax=6.6,ymin=-.025,ymax=1.075,xtick={-6,-3,0,3,6},minor xtick={-5,-4,-2,-1,1,2,4,5},
							ytick={0,.5,1},yticklabels={$0.0$,$0.5$,$1.0$},xlabel={$s$},yticklabels={},title={$E=0.35$ (dn)}]
						\addplot[paperink,line width=.45pt,no marks] coordinates {(-6.6,0) (6.6,0)};
						\addplot[ycomb,paperblue,line width=1.35pt,mark=*,mark size=1.85pt,mark options={solid,fill=paperblue}] coordinates {
							(-1,0.9185062129494437)
							(0,0.2819465736146959)
						};
						\addplot[ycomb,paperorange,line width=1.35pt,mark=*,mark size=1.85pt,mark options={solid,fill=paperorange}] coordinates {
							(-6,0.0003771943342965047)
							(-5,0.002139326460024926)
							(-4,0.01164823698281564)
							(-3,0.05945859042416979)
							(-2,0.2697842503332698)
							(1,0.01982877122959927)
							(2,0.0001257314720739471)
							(3,5.982128268716296e-07)
							(4,2.529967855151871e-09)
							(5,1.003097849568838e-11)
							(6,3.813553680954422e-14)
						};
						\node[anchor=south west,font=\PaperPanelFont] at (axis description cs:0,1.20) {(c)};
					\end{axis}
					\begin{axis}[paper plot,at={(14.875000000mm,102.240000000mm)},anchor=south west,width=46.59763314mm,height=25.135714286mm,
							xmin=-6.6,xmax=6.6,ymin=-.025,ymax=1.075,xtick={-6,-3,0,3,6},minor xtick={-5,-4,-2,-1,1,2,4,5},
							ytick={0,.5,1},yticklabels={$0.0$,$0.5$,$1.0$},xlabel={$s$},ylabel={$\qty|h_{s}|/\sqrt{2E}$},title={$E=0.49$ (dn)}]
						\addplot[paperink,line width=.45pt,no marks] coordinates {(-6.6,0) (6.6,0)};
						\addplot[ycomb,paperblue,line width=1.35pt,mark=*,mark size=1.85pt,mark options={solid,fill=paperblue}] coordinates {
							(-1,0.7217200219141506)
							(0,0.507431740475731)
						};
						\addplot[ycomb,paperorange,line width=1.35pt,mark=*,mark size=1.85pt,mark options={solid,fill=paperorange}] coordinates {
							(-6,0.009412516385870543)
							(-5,0.02669546819549073)
							(-4,0.07268429347929453)
							(-3,0.1855303902190487)
							(-2,0.4209793453102431)
							(1,0.06230783596711466)
							(2,0.003137679764720335)
							(3,0.0001193814366126832)
							(4,4.037720290936778e-06)
							(5,1.280286344888958e-07)
							(6,3.897168944058854e-09)
						};
						\node[anchor=south west,font=\PaperPanelFont] at (axis description cs:0,1.20) {(d)};
					\end{axis}
					\begin{axis}[paper plot,at={(70.326183430mm,102.240000000mm)},anchor=south west,width=46.59763314mm,height=25.135714286mm,
							xmin=-6.6,xmax=6.6,ymin=-.025,ymax=1.075,xtick={-6,-3,0,3,6},minor xtick={-5,-4,-2,-1,1,2,4,5},
							ytick={0,.5,1},yticklabels={$0.0$,$0.5$,$1.0$},xlabel={$s$},yticklabels={},title={$E=0.55$ (cn)}]
						\addplot[paperink,line width=.45pt,no marks] coordinates {(-6.6,0) (6.6,0)};
						\addplot[ycomb,paperblue,line width=1.35pt,mark=*,mark size=1.85pt,mark options={solid,fill=paperblue}] coordinates {
							(-1,0.3527883113334024)
							(0,0.370473450782172)
						};
						\addplot[ycomb,paperorange,line width=1.35pt,mark=*,mark size=1.85pt,mark options={solid,fill=paperorange}] coordinates {
							(-6,0.05874644000029011)
							(-5,0.1042692729509414)
							(-4,0.1782881698857994)
							(-3,0.2813081879530286)
							(-2,0.4245306027105748)
							(1,0.3527883113334027)
							(2,0.4245306027105747)
							(3,0.2813081879530286)
							(4,0.1782881698857993)
							(5,0.1042692729509413)
							(6,0.05874644000028997)
						};
						\node[anchor=south west,font=\PaperPanelFont] at (axis description cs:0,1.20) {(e)};
					\end{axis}
					\begin{axis}[paper plot,at={(125.777366860mm,102.240000000mm)},anchor=south west,width=46.59763314mm,height=25.135714286mm,
							xmin=-6.6,xmax=6.6,ymin=-.025,ymax=1.075,xtick={-6,-3,0,3,6},minor xtick={-5,-4,-2,-1,1,2,4,5},
							ytick={0,.5,1},yticklabels={$0.0$,$0.5$,$1.0$},xlabel={$s$},yticklabels={},title={$E=1.00$ (cn)}]
						\addplot[paperink,line width=.45pt,no marks] coordinates {(-6.6,0) (6.6,0)};
						\addplot[ycomb,paperblue,line width=1.35pt,mark=*,mark size=1.85pt,mark options={solid,fill=paperblue}] coordinates {
							(-1,0.5261818662827148)
							(0,0.04938460693423678)
						};
						\addplot[ycomb,paperorange,line width=1.35pt,mark=*,mark size=1.85pt,mark options={solid,fill=paperorange}] coordinates {
							(-6,0.01682952959321223)
							(-5,0.04085077281182672)
							(-4,0.09521481518424736)
							(-3,0.2076309878441491)
							(-2,0.4095498349119108)
							(1,0.5261818662827146)
							(2,0.4095498349119107)
							(3,0.2076309878441491)
							(4,0.09521481518424736)
							(5,0.04085077281182675)
							(6,0.01682952959321228)
						};
						\node[anchor=south west,font=\PaperPanelFont] at (axis description cs:0,1.20) {(f)};
					\end{axis}
					\begin{axis}[paper plot,at={(17.5mm,14.45mm)},anchor=south west,width=154.875mm,height=67.575mm,
							xmin=-.12,xmax=12.12,ymode=log,ymin=1e-30,ymax=10,xtick={0,1,...,12},ytick={1e-30,1e-25,1e-20,1e-15,1e-10,1e-5,1},
							minor y tick num=0,xlabel={$r$},ylabel={$\qty|\psi_{r,B}|^{2}/\qty|\psi_{0,B}|^{2}$},
							title={Edge profiles for (a)--(d) and the separatrix limit},title style={yshift=8pt},
							legend style={at={(.035,.035)},anchor=south west,legend columns=3,transpose legend,column sep=12pt,row sep=3pt,cells={anchor=west}}]
						\addplot[paperorange,only marks,mark=diamond*,mark size=2.25pt] coordinates {(0,1)};
						\addlegendentry{(a) Harmonic}
						\addplot[paperblue,mark=*,mark size=1.65pt,mark options={fill=white,line width=.65pt}] coordinates {
							(0,1)
							(1,0.01423283614753157)
							(2,0.0001144927885470445)
							(3,7.285788834689102e-07)
							(4,4.076848045830561e-09)
							(5,2.102899266403971e-11)
							(6,1.025422132062072e-13)
							(7,4.798596349536504e-16)
							(8,2.176063551853878e-18)
							(9,9.626227903580327e-21)
							(10,4.173723332324437e-23)
							(11,1.779886668083364e-25)
							(12,7.485406385421357e-28)
						};
						\addlegendentry{(b) $E=0.10$ (dn)}
						\addplot[paperteal,mark=square*,mark size=1.65pt,mark options={fill=white,line width=.65pt}] coordinates {
							(0,1)
							(1,0.08274124945278925)
							(2,0.003962530366654534)
							(3,0.000150999197590032)
							(4,5.071653768003455e-06)
							(5,1.572115315501234e-07)
							(6,4.610031401962245e-09)
							(7,1.29788148474474e-10)
							(8,3.541890138317078e-12)
							(9,9.430801164680397e-14)
							(10,2.461545229012965e-15)
							(11,6.319974731175132e-17)
							(12,1.600344447879605e-18)
						};
						\addlegendentry{(c) $E=0.35$ (dn)}
						\addplot[paperpurple,mark=triangle*,mark size=1.65pt,mark options={fill=white,line width=.65pt}] coordinates {
							(0,1)
							(1,0.2926229771138613)
							(2,0.05382641680387498)
							(3,0.008034983214284102)
							(4,0.00106576322211542)
							(5,0.0001310037919188441)
							(6,1.526949900595207e-05)
							(7,1.711308229973324e-06)
							(8,1.860960158407374e-07)
							(9,1.975906149529348e-08)
							(10,2.057620759271097e-09)
							(11,2.108536801996255e-10)
							(12,2.131658412123073e-11)
						};
						\addlegendentry{(d) $E=0.49$ (dn)}
						\addplot[paperink,densely dotted,line width=1.35pt,no marks] coordinates {(0,1) (12,1)};
						\addlegendentry{{$E\to\frac{1}{2}-0$ (fixed $r$)}}
						\addplot[paperorange,only marks,mark=diamond*,mark size=2.25pt,forget plot] coordinates {(0,1)};
						\node[anchor=south west,font=\PaperPanelFont] at (axis description cs:0,1.07) {(g)};
					\end{axis}
				\end{pgfinterruptboundingbox}
			\end{tikzpicture}
		\endgroup%
	}
	\caption{Changes in hopping amplitudes and edge-state profiles induced by nonlinearity.
		(a--f) Relative coupling amplitudes $\qty|h_{s}|/\sqrt{2E}$ on a linear scale.
		(a) For the harmonic oscillator, only $h_{-1}$ is nonzero. (b--d) The dn solutions in the left well at $E=0.10,0.35,0.49$.
		Blue denotes the nearest-neighbor couplings $s=-1,0$, and orange denotes the remaining long-range couplings.
		(e,f) The cn solutions spanning both wells at $E=0.55,1.00$, respectively.
		(g) Semilogarithmic plot of the zero-energy edge-mode profiles, rescaled so that the squared amplitude at the edge site $r=0$ is $1$. Solid lines show the dn solutions at $E=0.10,0.35,0.49$, corresponding to (b--d).
		For the harmonic oscillator in (a), only the point at $r=0$ is shown.
		The black dotted line shows the constant value $1$ obtained as $E\to1/2-0$ with each $r$ fixed.
		The zero-energy edge modes of the harmonic oscillator and the dn solutions are normalizable, whereas the squared norm of the dotted profile diverges.}
	\label{fig:nonlinear-hopping-edge}
\end{figure*}
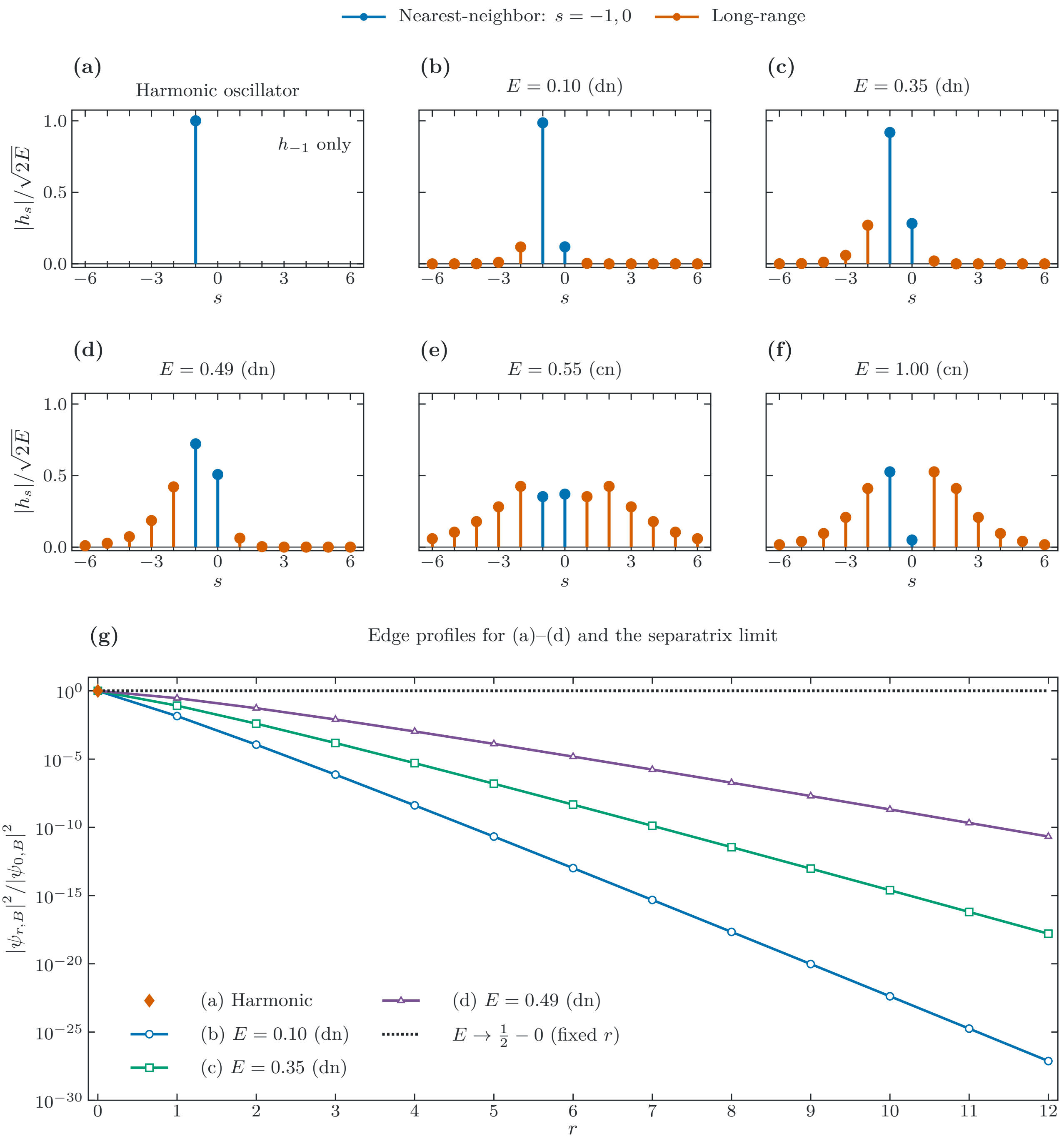

\section{Long-range hopping and edge states in the double well}
\label{sec:nonlinear-transitions}
\Jaindent Using the periodic orbits of the symmetric double well, we examine how anharmonicity and bifurcations of the motion affect the hopping amplitudes and edge states.

\subsection{Nonlinearity and longer-range hopping}
\Jaindent For small-amplitude motion in the left well, $h_{-1}$ dominates, as in the harmonic oscillator.
As the energy increases, $h_{-2},h_{-3},\ldots$ appear together with the intracell coupling $h_{0}$ [Fig.~\ref{fig:nonlinear-hopping-edge} (a--d)].
The variation in the rotation speed governed by $\dot Q=2\ii qQ$ generates higher harmonics and hence hopping to distant unit cells.

\Jaindent Above the barrier, the phase advances in both directions, and the distribution of couplings changes to that of the cn solution shown in Fig.~\ref{fig:nonlinear-hopping-edge} (e,f).
Both the dn and cn solutions satisfy Eq.~\eqref{eq:frequency-flatness-constraint}, so their bands are flat at $\pm\sqrt{2E}$.

\subsection{Separatrix and transition without gap closing}
\label{subsec:separatrix-transition}
\Jaindent As the separatrix is approached, the range of harmonics required to represent the motion increases, and the exponential decay of the hopping amplitudes becomes slower.
Throughout this approach, the Lax eigenvalues $\pm\sqrt{2E}$ approach $\pm1$, and the band gap around zero remains finite.
At $E=1/2$, however, the period diverges, and the closed curve over one period on which the winding number is defined is lost.
The intrawell periodic orbits and the orbit spanning both wells on opposite sides of the separatrix have different winding numbers, even though the band gap remains finite.
As the period diverges, the phase rotations of $Q$ occupy a vanishing fraction of each period. When a period is rescaled to a fixed interval, convergence to the limiting waveform is nonuniform, so the two families of periodic orbits are separated by a singular limit of the closed curve defining the winding number.

\Jaindent Topological transitions without bulk gap closing have also been found in models with algebraically decaying long-range couplings~\cite{Vodola2016,LeporiDellAnna2017}.
In the present model, the hopping amplitudes decay exponentially at every finite period. Their decay length diverges at the classical separatrix, linking the singular limit of the frequency chain to the bifurcation of the underlying motion.

\subsection{Edge states near the separatrix}
\Jaindent We next consider a semi-infinite chain obtained by keeping the same unit cell, retaining $r\ge0$, and removing all couplings across the boundary.
Its edge states are eigenstates localized in the frequency direction, and their profiles specify the weight on each harmonic.
The bulk--edge correspondence for chiral-symmetric systems~\cite{RyuHatsugai2002,Jones2023} and the edge equations in Appendix~\ref{app:double-well-edge} imply that the dn solution has one zero-energy edge mode.
The mode has support on sublattice B for the left well with $\nu=-1$ and on sublattice A for the right well with $\nu=+1$.
For the cn solution, the zero-eigenvalue equations in the same appendix confirm the disappearance of the zero-energy edge mode.

\Jaindent The normalized zero-energy edge mode $\psi_{r,B}$ for the left well approaches $\psi_{r,B}=\delta_{r,0}$ in the harmonic limit $E\to+0$ and spreads through long-range hopping as the energy increases.
We denote the exponential decay length of its amplitude in harmonic-index units by $\xi\qty(E)$ (an explicit expression is given in Appendix~\ref{app:double-well-edge}).

\Jaindent As $E\to1/2-0$,
\begin{align}
	\xi\qty(E)\simeq\frac{1}{\pi^{2}}
	\ln\qty(\frac{32}{1-\sqrt{2E}}),
	\label{eq:double-well-edge-limit}
\end{align}
so the localization length in the frequency lattice diverges.

\Jaindent Figure~\ref{fig:nonlinear-hopping-edge} (g) shows the relative squared amplitude $\qty|\psi_{r,B}|^{2}/\qty|\psi_{0,B}|^{2}$ of the zero-energy edge mode for the harmonic oscillator and the dn solutions in (b--d).
The zero-energy edge mode spreads as the energy increases, and the relative squared amplitude approaches $1$ at each $r$ as $E\to1/2-0$.
The constant profile obtained in the critical limit has a divergent squared norm, and the normalizable zero-energy edge mode is lost.
\section{Supersymmetric quantization and the Witten index}
\label{sec:susy-witten}
\Jaindent The Lax operator used in this paper yields a frequency chain when evaluated along a periodic orbit and a supercharge of a supersymmetric quantum mechanical system when canonically quantized.
The mode-shell correspondence relates chiral zero-mode indices to winding numbers of operator symbols in phase space in a semiclassical limit~\cite{JezequelDelplace2024ModeShell}.
For the present one-degree-of-freedom system, we give an explicit decomposition of the Witten index into the winding numbers of classical periodic orbits at a common positive energy.
We consider bound systems on the real line and take $w$ to be a smooth real function independent of $\hbar$, satisfying $w\qty(q)^{2}\to\infty$ as $\abs{q}\to\infty$.
For the quantum system, we assume a discrete spectrum of bound states with a finite thermal trace.

\subsection{Supercharge and fermion parity}
\Jaindent Following the construction of supersymmetric quantum mechanics~\cite{Cooper1995}, we canonically quantize the Lax operator $\mathsf L=p\sigma^{x}+w\qty(q)\sigma^{y}$ before substituting a periodic solution and construct a supercharge acting on two-component wave functions.
We define the canonical variables and the supercharge as
\begin{align}
	&\hat q=q,\qquad \hat p=-\ii\hbar\dv{q},\qquad
	\comm{\hat q}{\hat p}=\ii\hbar,\notag\\
	&\hat R\coloneq\hat{\mathsf L}
	=\hat p\sigma^{x}+w\qty(q)\sigma^{y}.
	\label{eq:susy-quantized-lax}
\end{align}
Taking the upper component to be even and the lower component to be odd, we choose the fermion parity as $\qty(-1)^{F}=\sigma^{z}$.
Chiral symmetry of the Lax operator gives
\begin{align}
	&\acomm{\hat R}{\qty(-1)^{F}}=0,\qquad
	\hat H_{\mathrm{SUSY}}\coloneq\frac{1}{2}\hat R^{2},\notag\\
	&\comm{\hat H_{\mathrm{SUSY}}}{\qty(-1)^{F}}=0.
	\label{eq:susy-supercharges}
\end{align}
The supercharge $\hat R$ connects states of opposite parity, while $\hat H_{\mathrm{SUSY}}$ evolves states within each parity sector.
Using $\comm{\hat p}{w\qty(q)}=-\ii\hbar w'\qty(q)$, we obtain
\begin{align}
	&\hat H_{\mathrm{SUSY}}
	=\frac{1}{2}\qty[\hat p^{2}+w\qty(q)^{2}]\mathsf{I}
	+\frac{\hbar}{2}w'\qty(q)\sigma^{z}\notag\\
	&\phantom{\hat H_{\mathrm{SUSY}}}=\begin{pmatrix}\hat H_{+}&0\\0&\hat H_{-}\end{pmatrix},\notag\\
	&\hat H_{\pm}=\frac{1}{2}\qty[-\hbar^{2}\dv[2]{q}
		+w\qty(q)^{2}\pm\hbar w'\qty(q)].
	\label{eq:susy-partner-hamiltonians}
\end{align}
The two parity sectors share the classical Hamiltonian \eqref{eq:hamiltonian} and have quantum corrections of opposite signs.

\Jaindent The Witten index counts quantum states with a sign determined by their fermion parity~\cite{Witten1982Index}.
With our conventions,
\begin{align}
	&\Delta_{\mathrm W}\coloneq\operatorname{Tr}\qty[
		\qty(-1)^{F}\exp\qty(-\beta\hat H_{\mathrm{SUSY}})]\notag\\
	&\phantom{\Delta_{\mathrm W}}=n_{+}\qty(0)-n_{-}\qty(0),\qquad\beta>0.
	\label{eq:witten-index}
\end{align}
Here, $n_{\pm}\qty(0)$ counts the normalizable zero-energy states of $\hat H_{\pm}$.
The supercharge $\hat R$ pairs positive-energy states with states of the same energy and opposite parity, so their contributions to the index cancel.
Thus, $\Delta_{\mathrm W}$ is determined by the difference between the numbers of zero-energy states of the two parities that are normalizable over the full space.

\Jaindent Factorization of $\hat H_{\pm}$ shows that the zero-energy wave functions satisfy
\begin{align}
	&\qty[\hbar\dv{q}\mp w\qty(q)]\psi_{\pm}\qty(q)=0,\notag\\
	&\psi_{\pm}\qty(q)\propto
	\exp\qty[\pm\frac{1}{\hbar}\int_{0}^{q}\dd{q'}\,w\qty(q')].
	\label{eq:susy-zero-wavefunctions}
\end{align}
Counting the wave functions that decay at both infinities~\cite{Cooper1995} gives
\begin{align}
	\Delta_{\mathrm W}=
	\frac{\operatorname{sgn}\qty[w\qty(-\infty)]
		-\operatorname{sgn}\qty[w\qty(+\infty)]}{2}.
	\label{eq:witten-asymptotic-signs}
\end{align}

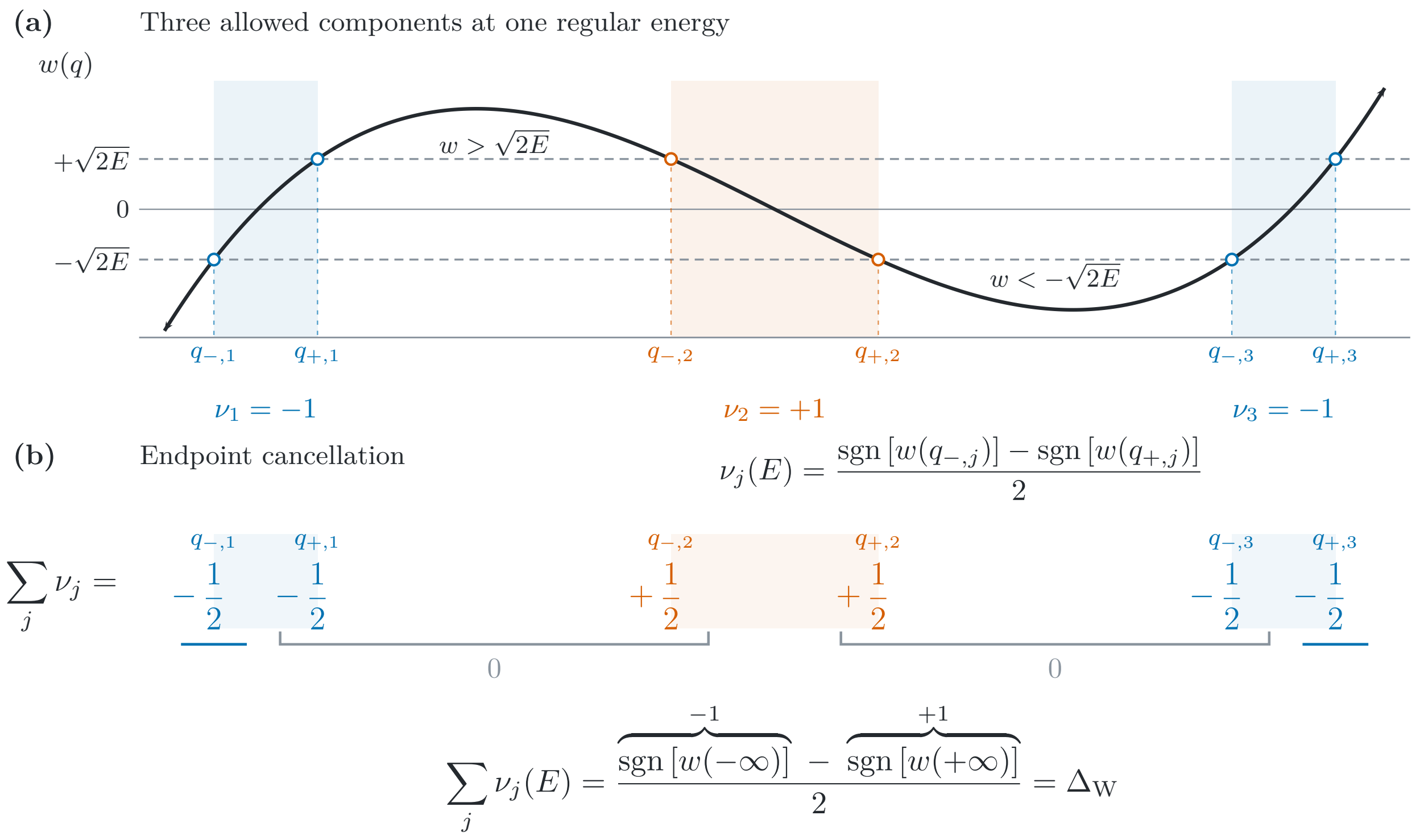
\begin{figure*}[!t]
	\centering
	\resizebox{175mm}{!}{%
		\begingroup%
			\begin{tikzpicture}[x=1mm,y=1mm,every node/.style={font=\PaperLabelFont,text=paperink,inner sep=0pt}]
				\path[use as bounding box] (0,0) rectangle (175,100);
				\begin{pgfinterruptboundingbox}
					\node[anchor=west,font=\PaperPanelFont] at (6.125,96.2) {(a)};
					\node[anchor=west,font=\fontsize{9}{10.8}\selectfont] at (21,96.2) {Three allowed components at one regular energy};
					\begin{scope}[shift={(95.375,74.5)},x=34.91784038mm,y=5.88235294mm]
						\fill[paperblue,opacity=.08] (-1.87938524157,-2.55) rectangle (-1.53208888624,2.55);
						\fill[paperorange,opacity=.08] (-0.347296355334,-2.55) rectangle (0.347296355334,2.55);
						\fill[paperblue,opacity=.08] (1.53208888624,-2.55) rectangle (1.87938524157,2.55);
						\draw[paperguide,line width=.65pt,dash pattern=on 3pt off 2pt] (-2.13,1)--(2.13,1);
						\draw[paperguide,line width=.65pt,dash pattern=on 3pt off 2pt] (-2.13,-1)--(2.13,-1);
						\draw[paperguide,line width=.45pt] (-2.13,0)--(2.13,0);
						\draw[paperink,line width=1.2pt] plot[domain=-2.045:2.045,samples=401] (\x,{\x*\x*\x-3*\x});
						\draw[paperblue,opacity=.65,line width=.45pt,dash pattern=on 1.4pt off 2pt] (-1.87938524157,-2.5)--(-1.87938524157,-1);
						\draw[paperblue,fill=white,line width=.8pt] (-1.87938524157,-1) circle[radius=1.9pt];
						\draw[paperblue,opacity=.65,line width=.45pt,dash pattern=on 1.4pt off 2pt] (-1.53208888624,-2.5)--(-1.53208888624,1);
						\draw[paperblue,fill=white,line width=.8pt] (-1.53208888624,1) circle[radius=1.9pt];
						\draw[paperorange,opacity=.65,line width=.45pt,dash pattern=on 1.4pt off 2pt] (-0.347296355334,-2.5)--(-0.347296355334,1);
						\draw[paperorange,fill=white,line width=.8pt] (-0.347296355334,1) circle[radius=1.9pt];
						\draw[paperorange,opacity=.65,line width=.45pt,dash pattern=on 1.4pt off 2pt] (0.347296355334,-2.5)--(0.347296355334,-1);
						\draw[paperorange,fill=white,line width=.8pt] (0.347296355334,-1) circle[radius=1.9pt];
						\draw[paperblue,opacity=.65,line width=.45pt,dash pattern=on 1.4pt off 2pt] (1.53208888624,-2.5)--(1.53208888624,-1);
						\draw[paperblue,fill=white,line width=.8pt] (1.53208888624,-1) circle[radius=1.9pt];
						\draw[paperblue,opacity=.65,line width=.45pt,dash pattern=on 1.4pt off 2pt] (1.87938524157,-2.5)--(1.87938524157,1);
						\draw[paperblue,fill=white,line width=.8pt] (1.87938524157,1) circle[radius=1.9pt];
						\draw[paperink,line width=.8pt,-{Stealth[length=3pt,width=2.5pt]}] (-2.012,-2.108865728)--(-2.045,-2.417241125);
						\draw[paperink,line width=.8pt,-{Stealth[length=3pt,width=2.5pt]}] (2.012,2.108865728)--(2.045,2.417241125);
						\node[anchor=south] at (-0.939692620786,1.05) {$w>\sqrt{2E}$};
						\node[anchor=north] at (0.939692620786,-1.05) {$w<-\sqrt{2E}$};
					\end{scope}
					\draw[paperguide,line width=.55pt] (21,59.5)--(169.75,59.5);
					\node[anchor=south east,font=\fontsize{9}{10.8}\selectfont] at (15.79375,89.65) {$w\qty(q)$};
					\node[anchor=north west,font=\fontsize{9}{10.8}\selectfont] at (171.535,58.15) {$q$};
					\node[anchor=east] at (19.94,68.6176470588) {$-\sqrt{2E}$};
					\node[anchor=east] at (19.94,74.5) {$0$};
					\node[anchor=east] at (19.94,80.3823529412) {$+\sqrt{2E}$};
					\node[anchor=north,text=paperblue,font=\fontsize{8.3}{9.96}\selectfont] at (29.7509261306,58.44) {$q_{-,1}$};
					\node[anchor=north,text=paperblue,font=\fontsize{8.3}{9.96}\selectfont] at (41.8777648291,58.44) {$q_{+,1}$};
					\node[anchor=north,text=paperorange,font=\fontsize{8.3}{9.96}\selectfont] at (83.2481613014,58.44) {$q_{-,2}$};
					\node[anchor=north,text=paperorange,font=\fontsize{8.3}{9.96}\selectfont] at (107.501838699,58.44) {$q_{+,2}$};
					\node[anchor=north,text=paperblue,font=\fontsize{8.3}{9.96}\selectfont] at (148.872235171,58.44) {$q_{-,3}$};
					\node[anchor=north,text=paperblue,font=\fontsize{8.3}{9.96}\selectfont] at (160.999073869,58.44) {$q_{+,3}$};
					\node[anchor=north,text=paperblue,font=\fontsize{9.4}{11.28}\selectfont] at (35.8143454798,52.3) {$\nu_{1}=-1$};
					\fill[paperblue,opacity=.06] (29.7509261306,25.5) rectangle (41.8777648291,36.5);
					\node[anchor=north,text=paperorange,font=\fontsize{9.4}{11.28}\selectfont] at (95.375,52.3) {$\nu_{2}=+1$};
					\fill[paperorange,opacity=.06] (83.2481613014,25.5) rectangle (107.501838699,36.5);
					\node[anchor=north,text=paperblue,font=\fontsize{9.4}{11.28}\selectfont] at (154.93565452,52.3) {$\nu_{3}=-1$};
					\fill[paperblue,opacity=.06] (148.872235171,25.5) rectangle (160.999073869,36.5);
					\node[anchor=west,font=\PaperPanelFont] at (6.125,45.5) {(b)};
					\node[anchor=west,font=\fontsize{9}{10.8}\selectfont] at (21,45.5) {Endpoint cancellation};
					\node[font=\fontsize{10}{12}\selectfont] at (117.25,44.2) {$\displaystyle\nu_{j}\qty(E)=\frac{\operatorname{sgn}\qty[w\qty(q_{-,j})]-\operatorname{sgn}\qty[w\qty(q_{+,j})]}{2}$};
					\node[font=\fontsize{11}{13.2}\selectfont] at (11.9,29.2) {$\displaystyle\sum_{j}\nu_{j}=$};
					\node[text=paperblue,font=\fontsize{8.2}{9.84}\selectfont] at (29.7509261306,35.5) {$q_{-,1}$};
					\node[text=paperblue,font=\fontsize{11}{13.2}\selectfont] (half0) at (29.7509261306,29.2) {$\displaystyle\frac{1}{2}$};
					\node[anchor=east,text=paperblue,font=\fontsize{11}{13.2}\selectfont,xshift=-1.4pt] at (half0.west) {$-$};
					\node[text=paperblue,font=\fontsize{8.2}{9.84}\selectfont] at (41.8777648291,35.5) {$q_{+,1}$};
					\node[text=paperblue,font=\fontsize{11}{13.2}\selectfont] (half1) at (41.8777648291,29.2) {$\displaystyle\frac{1}{2}$};
					\node[anchor=east,text=paperblue,font=\fontsize{11}{13.2}\selectfont,xshift=-1.4pt] at (half1.west) {$-$};
					\node[text=paperorange,font=\fontsize{8.2}{9.84}\selectfont] at (83.2481613014,35.5) {$q_{-,2}$};
					\node[text=paperorange,font=\fontsize{11}{13.2}\selectfont] (half2) at (83.2481613014,29.2) {$\displaystyle\frac{1}{2}$};
					\node[anchor=east,text=paperorange,font=\fontsize{11}{13.2}\selectfont,xshift=-1.4pt] at (half2.west) {$+$};
					\node[text=paperorange,font=\fontsize{8.2}{9.84}\selectfont] at (107.501838699,35.5) {$q_{+,2}$};
					\node[text=paperorange,font=\fontsize{11}{13.2}\selectfont] (half3) at (107.501838699,29.2) {$\displaystyle\frac{1}{2}$};
					\node[anchor=east,text=paperorange,font=\fontsize{11}{13.2}\selectfont,xshift=-1.4pt] at (half3.west) {$+$};
					\node[text=paperblue,font=\fontsize{8.2}{9.84}\selectfont] at (148.872235171,35.5) {$q_{-,3}$};
					\node[text=paperblue,font=\fontsize{11}{13.2}\selectfont] (half4) at (148.872235171,29.2) {$\displaystyle\frac{1}{2}$};
					\node[anchor=east,text=paperblue,font=\fontsize{11}{13.2}\selectfont,xshift=-1.4pt] at (half4.west) {$-$};
					\node[text=paperblue,font=\fontsize{8.2}{9.84}\selectfont] at (160.999073869,35.5) {$q_{+,3}$};
					\node[text=paperblue,font=\fontsize{11}{13.2}\selectfont] (half5) at (160.999073869,29.2) {$\displaystyle\frac{1}{2}$};
					\node[anchor=east,text=paperblue,font=\fontsize{11}{13.2}\selectfont,xshift=-1.4pt] at (half5.west) {$-$};
					\draw[paperguide,line width=.85pt] (37.5027648291,25.1)--(37.5027648291,23.6)--(87.6231613014,23.6)--(87.6231613014,25.1);
					\node[anchor=north,text=paperguide,font=\fontsize{9.5}{11.4}\selectfont] at (62.5629630653,21.9) {$0$};
					\draw[paperguide,line width=.85pt] (103.126838699,25.1)--(103.126838699,23.6)--(153.247235171,23.6)--(153.247235171,25.1);
					\node[anchor=north,text=paperguide,font=\fontsize{9.5}{11.4}\selectfont] at (128.187036935,21.9) {$0$};
					\draw[paperblue,line width=.9pt] (25.9009261306,23.6)--(33.6009261306,23.6);
					\draw[paperblue,line width=.9pt] (157.149073869,23.6)--(164.849073869,23.6);
					\node[font=\fontsize{10.5}{12.6}\selectfont] at (96.25,9) {$\displaystyle\sum_{j}\nu_{j}\qty(E)=\frac{\overbrace{\operatorname{sgn}\qty[w\qty(-\infty)]}^{-1}\ -\ \overbrace{\operatorname{sgn}\qty[w\qty(+\infty)]}^{+1}}{2}=\Delta_{\mathrm{W}}$};
				\end{pgfinterruptboundingbox}
			\end{tikzpicture}
		\endgroup%
	}
	\caption{Correspondence between the sum of winding numbers obtained from the endpoint formula and the Witten index.
		(a) Schematic $w\qty(q)$ with three periodic orbits at a common energy $E>0$. Horizontal dashed lines denote $\pm\sqrt{2E}$, and colored bands mark the allowed regions $w\qty(q)^{2}<2E$. The signs at the turning points $q_{-,j},q_{+,j}$ of each region give $\nu_{1}=-1$, $\nu_{2}=+1$, and $\nu_{3}=-1$.
		(b) Contributions from the individual endpoints, shown directly below (a) in the same colors. The interior contributions joined by gray brackets cancel, leaving the two terms underlined in blue. The formula in the bottom row, expressed in terms of the signs at infinity, gives $\sum_{j}\nu_{j}=-1=\Delta_{\mathrm W}$ for this example.}
	\label{fig:local-winding-witten}
\end{figure*}

\subsection{From local winding numbers to the global index}
\Jaindent We consider all periodic orbits at a common energy $E>0$ in the classical system with the same $w$.
We choose an energy for which $w'\qty(q)\ne0$ at every turning point and the fundamental period of each orbit is finite.
We number the allowed regions $w\qty(q)^{2}<2E$ from left to right and count one periodic orbit traversing each region back and forth.
With the left and right turning points denoted by $q_{-,j},q_{+,j}$, Eq.~\eqref{eq:winding-turning-points} gives
\begin{align}
	\nu_{j}\qty(E)=
	\frac{\operatorname{sgn}\qty[w\qty(q_{-,j})]
		-\operatorname{sgn}\qty[w\qty(q_{+,j})]}{2}.
	\label{eq:winding-each-component}
\end{align}
Between adjacent allowed regions, $w\qty(q)^{2}>2E$, so the sign of $w$ remains unchanged.
The contributions from $q_{+,j}$ and $q_{-,j+1}$ therefore have opposite signs and cancel.
The sign also remains unchanged from each outermost turning point to infinity, leaving only the signs of $w\qty(-\infty)$ and $w\qty(+\infty)$ in the sum.
Combining this result with Eq.~\eqref{eq:witten-asymptotic-signs}, we obtain
\begin{align}
	\sum_{j}\nu_{j}\qty(E)
	&=\frac{\operatorname{sgn}\qty[w\qty(-\infty)]
		-\operatorname{sgn}\qty[w\qty(+\infty)]}{2}\notag\\
	&=\Delta_{\mathrm W}.
	\label{eq:witten-winding-sum}
\end{align}
This cancellation of endpoint contributions is illustrated in Fig.~\ref{fig:local-winding-witten}.
Changing the classical energy can change the number of allowed regions and the winding number of each periodic orbit, but their sum gives the same Witten index of the quantum system.
For a periodic orbit approaching a separatrix, the period diverges and its winding number becomes undefined in the limit.
Equation~\eqref{eq:witten-winding-sum} holds for the finite-period solutions on both sides of the bifurcation.

\Jaindent The correspondence between the classical phase and the Witten index for a single pair of turning points has been discussed in semiclassical quantization of supersymmetric quantum mechanics~\cite{Junker1995}.
With our conventions, the winding number of the single periodic orbit gives
\begin{align}
	\Delta_{\mathrm W}=\nu\qty(E).
	\label{eq:witten-high-energy-winding}
\end{align}
Equation~\eqref{eq:witten-winding-sum} expresses the same index as a sum over all allowed regions at a common energy.
Each term is the winding number of the frequency chain generated by the corresponding periodic orbit, and the cancellation of interior turning-point contributions preserves the total across bifurcations.

\subsection{Semiclassical states and tunneling}
\Jaindent In the semiclassical approximation, the quantum corrections $\pm\hbar w'\qty(q)/2$ in $\hat H_{\pm}$ contribute to the phase accumulated under time evolution along a classical periodic orbit~\cite{InomataJunker1994}.
For the $j$th periodic orbit, the integral determining this phase follows from Eq.~\eqref{eq:winding-potential} as
\begin{align}
	\frac{1}{2}\int_{0}^{T_{j}}\dd{t}\,w'\qty(q\qty(t))
	=-\pi\nu_{j}\qty(E).
	\label{eq:susy-semiclassical-phase}
\end{align}
The phase due to the quantum correction over one period is therefore $+\pi\nu_{j}$ for $\hat H_{+}$ and $-\pi\nu_{j}$ for $\hat H_{-}$, and is determined by the winding number of the classical periodic orbit.

\Jaindent In particular, if $w$ has finitely many zeros $q_{j}$ with $w'\qty(q_{j})\ne0$, the periodic orbits as $E\to+0$ separate into orbits around the individual zeros.
The turning-point formula gives
\begin{align}
	\lim_{E\to+0}\nu_{j}\qty(E)
	=-\operatorname{sgn}\qty[w'\qty(q_{j})].
	\label{eq:low-energy-winding}
\end{align}
Consequently,
\begin{align}
	\Delta_{\mathrm W}
	=\sum_{j}\lim_{E\to+0}\nu_{j}\qty(E)
	=-\sum_{j}\operatorname{sgn}\qty[w'\qty(q_{j})].
	\label{eq:witten-low-energy-sum}
\end{align}
The correspondence between a signed count of classical zero-energy configurations and the Witten index has also been studied~\cite{Lo2022}.

\Jaindent The relation between this sign and the parity of the quantum state follows from the harmonic approximation in each well~\cite{Witten1982Morse}.
Approximating $w\qty(q)\simeq w'\qty(q_{j})\qty(q-q_{j})$ near $q_{j}$ gives the local energy levels of $\hat H_{\pm}$ as
\begin{align}
	&\varepsilon_{n,\pm}^{j}\simeq
	\hbar\abs{w'\qty(q_{j})}\qty(n+\frac{1}{2})
	\pm\frac{\hbar}{2}w'\qty(q_{j}),\notag\\
	&n=0,1,2,\ldots.
	\label{eq:susy-local-harmonic-spectrum}
\end{align}
The zero-point energy is canceled in $\hat H_{-}$ for $w'\qty(q_{j})>0$ and in $\hat H_{+}$ for $w'\qty(q_{j})<0$.
The corresponding local states have parity $-1,+1$, respectively, matching the winding numbers of the small-amplitude classical periodic orbits.

\Jaindent For the harmonic oscillator $w\qty(q)=q$, $\hat H_{-}$ has one zero-energy state, and $\Delta_{\mathrm W}=-1$ agrees with $\nu=-1$ at any $E>0$.
For the symmetric double well $w\qty(q)=1-q^{2}$, the winding numbers $-1,+1$ of the left and right wells sum to zero for $0<E<1/2$, while the single periodic orbit spanning both wells has winding number zero for $E>1/2$.
Both candidate quantum zero-energy wave functions $\exp\qty[\pm\qty(q-q^{3}/3)/\hbar]$ diverge at one of the two infinities, so no normalizable zero-energy state exists and $\Delta_{\mathrm W}=0$.

\Jaindent For small $\hbar$ in the double well, the left well supports a local state with odd parity and the right well a local state with even parity, each with zero energy in its harmonic approximation.
Overlap of the wave-function tails extending into the barrier produces a small matrix element of the supercharge $\hat R$ connecting the two local states~\cite{Witten1982Morse}.
In the low-energy two-state approximation, $\hat H_{\mathrm{SUSY}}=\hat R^{2}/2$ implies that half the squared modulus of this matrix element gives the small positive energy shared by both states.
The two states form a SUSY doublet, and their contributions to the Witten index cancel.
This tunneling picture connects the local cancellation of zero-point energy in each well to the existence or absence of zero-energy states normalizable over the full space.

\Jaindent Thus, the winding number of a small-amplitude periodic orbit in each well corresponds to the parity of a local quantum state.
Summing these winding numbers gives the Witten index of the full system, including tunneling between wells.

\section{Conclusion}
\label{sec:conclusion}
{\color{black}
	\Jaindent We have constructed an exact chiral-symmetric flat-band model in frequency space from nonlinear periodic orbits of Hamiltonian systems with one degree of freedom.
	The Fourier coefficients of the time-periodic Lax operator become hopping amplitudes between harmonics, and higher harmonics of the nonlinear waveform generate long-range couplings.
	Energy conservation imposes cancellation constraints on the hopping amplitudes and keeps both bands flat even in the presence of long-range couplings.
	The harmonic oscillator yields a fully dimerized topological SSH model, and nonlinearity extends this topological phase to models with long-range hopping.

	\Jaindent In the symmetric double well, the bifurcation from intrawell periodic orbits to an orbit spanning both wells corresponds to a change in the winding number and the disappearance of the zero-energy edge mode in the semi-infinite chain with a fixed boundary termination.
	The edge-mode localization length, measured in harmonic-index units, diverges logarithmically as the intrawell orbit approaches the separatrix, while the band gap remains finite.
	At the separatrix, the period diverges and the finite-period winding number becomes undefined; the periodic orbits on its two sides have different winding numbers.

	\Jaindent Canonical quantization connects this classical topology to quantum supersymmetry.
	For bound motion on the real line with simple turning points, the chiral winding numbers of all periodic orbits at a common positive energy sum exactly to the Witten index.
	This decomposes the quantum index into contributions from classical periodic orbits, each of which also defines the winding number of a frequency chain.

	\Jaindent Traveling-wave reductions extend the construction to periodic nonlinear waves.
	For the nonlinear Schr\"odinger (NLS) and modified Korteweg--de Vries (mKdV) equations, nodeless dn wave trains have winding numbers $\nu=\pm1$, whereas sign-alternating cn wave trains have $\nu=0$.
	In their solitary-wave limits, the period diverges and the finite-period winding number becomes undefined.

	\Jaindent A natural extension is to more general classical integrable systems, such as tops and finite Toda lattices.
	For systems with a Hermitian Lax representation and periodic orbits, this construction yields flat bands in frequency space; their symmetries, topological invariants, and boundary states remain to be investigated.

	\Jaindent For nonlinear waves, studies of topology based on Lax eigenstates of the periodic Toda lattice~\cite{SatoFukui2023} could be extended to periodic soliton solutions in continuum systems.
	The Lax operators of continuum wave equations could be used to investigate the conditions for defining Chern numbers and their correspondence with boundary states.
	Further questions include constructing invariants that incorporate solitary-wave boundary conditions and clarifying their relations to soliton charges and localized modes.

	\Jaindent For the quantum correspondence, the exact decomposition of the Witten index provides a basis for further semiclassical quantization and tunneling calculations.
	These calculations could determine how the local states characterized by the individual-well winding numbers contribute to the low-energy spectrum.
	Quasiperiodic solutions with multiple fundamental frequencies offer a route to higher-dimensional frequency lattices in systems with multiple degrees of freedom.
	Studying weak departures from integrability could then reveal how the frequency lattice, band flatness, and edge states change as Kolmogorov--Arnold--Moser tori deform or break down and chaotic motion emerges.
	\par}

\begin{acknowledgments}
	For valuable discussions, we thank \mbox{M.~Kawahira}, \mbox{M.~Sato}, \mbox{K.~Shiozaki}, \mbox{K.~Totsuka}, and \mbox{X.~Wang}.
	U.M. acknowledges support from a Research Fellowship for Young Scientists (DC2) of the Japan Society for the Promotion of Science (JSPS) and JSPS KAKENHI Grant Number~JP25KJ1573.
	H.H. acknowledges partial support from JSPS KAKENHI Grant Number~JP26K06960.
\end{acknowledgments}

\appendix

\section{Derivation of the turning-point formula}
\label{app:details}
\label{app:turning-point-winding}
\Jaindent We derive the turning-point formula \eqref{eq:winding-turning-points}.
For a periodic orbit with energy $E>0$, energy conservation \eqref{eq:conserved-energy} gives
\begin{align}
	p=\pm\sqrt{2E-w\qty(q)^{2}}.
	\label{eq:turning-point-momentum}
\end{align}
Let $q_{-}<q_{+}$ be two adjacent simple turning points, with $w\qty(q_{\pm})^{2}=2E$ and $w\qty(q_{\pm})w'\qty(q_{\pm})\ne0$.
In the interior of this interval, $2E-w\qty(q)^{2}>0$, and the periodic orbit traversing the interval back and forth has a finite period.

\Jaindent Taking $p>0$ on the forward path and $p<0$ on the return path and using $\dd{t}=\dd{q}/p$, we obtain
\begin{equation}
	\int_{0}^{T}\dd{t}\,w'\qty(q\qty(t))
	=2\int_{q_{-}}^{q_{+}}\dd{q}\,\frac{w'\qty(q)}{\sqrt{2E-w\qty(q)^{2}}}.
	\label{eq:turning-point-time-integral}
\end{equation}
Substituting this result into Eq.~\eqref{eq:winding-potential} gives
\begin{align}
	\nu&=-\frac{1}{\pi}\int_{q_{-}}^{q_{+}}\dd{q}\,
	\frac{w'\qty(q)}{\sqrt{2E-w\qty(q)^{2}}}\notag\\
	&=-\frac{1}{\pi}\eval{\arcsin\qty(\frac{w\qty(q)}{\sqrt{2E}})}_{q=q_{-}}^{q=q_{+}}\notag\\
	&=\frac{\operatorname{sgn}\qty[w\qty(q_{-})]-\operatorname{sgn}\qty[w\qty(q_{+})]}{2}.
	\label{eq:turning-point-derivation}
\end{align}
Here, we use the principal branch of $\arcsin$ with values in $\qty[-\pi/2,\pi/2]$.

\section{Flatness constraints}
\label{app:frequency-flatness}
\Jaindent Substituting the Fourier expansion \eqref{eq:frequency-fourier} into the conservation law $\qty|Q|^{2}=2E$ gives
\begin{align}
	&\sum_{r\in\Z}h_{r+s}h_{r}^{*}
	=\frac{1}{T}\int_{0}^{T}\dd{t}\,
	\e^{-\ii s\Omega_{0}t}\qty|Q\qty(t)|^{2}\notag\\
	&\phantom{\sum_{r\in\Z}h_{r+s}h_{r}^{*}}=2E\delta_{s,0},
\end{align}
which yields Eq.~\eqref{eq:frequency-flatness-constraint}.

\Jaindent Suppose that only finitely many $h_{r}$ are nonzero, with smallest and largest indices $a<b$. For $s=b-a$, we have
\begin{align}
	&\sum_{r\in\Z}h_{r+b-a}h_{r}^{*}=h_{b}h_{a}^{*}\ne0.
\end{align}
This contradicts Eq.~\eqref{eq:frequency-flatness-constraint}, whose right-hand side vanishes.
Thus, for a chiral-symmetric two-band model with one component per sublattice, finite-range hopping and exact flatness are compatible only when a single Fourier coefficient is nonzero.

\Jaindent Truncating the hopping to a finite range makes the cancellation incomplete and generally produces dispersion.
In the semi-infinite chain considered in the main text, all hopping terms between the retained sites are kept.

\section{Explicit periodic solutions and winding numbers}
\label{app:periodic-solutions}
\Jaindent Table~\ref{tab:periodic-orbits} summarizes the relation between turning-point signs and winding numbers for the symmetric double well, power-law oscillators, and the pendulum.
We denote the parameter of the Jacobi elliptic functions by $m$ and define the complete elliptic integral as
\begin{align}
	K\qty(m)=\int_{0}^{\pi/2}\dd{x}\,
	\frac{1}{\sqrt{1-m\sin\qty(x)^{2}}}.
	\label{eq:elliptic-integral-definition}
\end{align}
The momentum of each solution follows from $p=\dot{q}$.

\subsection{Symmetric double well}
\Jaindent For $w\qty(q)=1-q^{2}$, energy conservation gives
\begin{align}
	&\dot q^{2}+\qty(1-q^{2})^{2}=2E.
	\label{eq:symmetric-scaled-energy}
\end{align}
With the outer turning point chosen at $t=0$, the intrawell periodic orbits for $0<E<1/2$ are
\begin{align}
	&q\qty(t)=\pm\sqrt{1+\sqrt{2E}}\operatorname{dn}\qty(
	\sqrt{1+\sqrt{2E}}\,t\,\middle|\,
	\frac{2\sqrt{2E}}{1+\sqrt{2E}}),\notag\\
	&T=\frac{2K\qty(\dfrac{2\sqrt{2E}}{1+\sqrt{2E}})}
	{\sqrt{1+\sqrt{2E}}}.
	\label{eq:symmetric-dn-solutions}
\end{align}
The minus and plus signs correspond to the left and right wells, respectively.
For $E>1/2$, the periodic orbit spanning both wells is
\begin{align}
	&q\qty(t)=\sqrt{1+\sqrt{2E}}\operatorname{cn}\qty(
	\sqrt{2\sqrt{2E}}\,t\,\middle|\,
	\frac{1+\sqrt{2E}}{2\sqrt{2E}}),\notag\\
	&T=\frac{4K\qty(\dfrac{1+\sqrt{2E}}{2\sqrt{2E}})}
	{\sqrt{2\sqrt{2E}}}.
	\label{eq:symmetric-cn-solution}
\end{align}
Here, the right turning point is chosen as the time origin.
For both the dn solution as $E\to1/2-0$ and the cn solution as $E\to1/2+0$, the elliptic parameter approaches $1$, and $K$ and the period diverge.

\subsection{Power-law oscillators}
\label{subsec:power-law}
\Jaindent Let $w\qty(q)=q^{n}$ for a positive integer $n$.
The case $n=1$ corresponds to the harmonic oscillator, while $n\geq2$ gives nonlinear oscillators.
The turning points are $q_{\pm}=\pm\qty(2E)^{1/\qty(2n)}$, and
\begin{align*}
	w\qty(q_{-})=\qty(-1)^{n}\sqrt{2E},\qquad
	w\qty(q_{+})=\sqrt{2E}.
\end{align*}
The turning-point formula \eqref{eq:winding-turning-points} therefore gives
\begin{align}
	\nu=\frac{\qty(-1)^{n}-1}{2}
	=\begin{cases}
		-1, & n\text{ odd},\\
		0, & n\text{ even}
	\end{cases}.
	\label{eq:power-law-winding}
\end{align}
For odd $n$, $Q$ makes a full turn around the circle. For even $n$, $w>0$ at both turning points, and $Q$ traverses an arc back and forth.
This distinction holds for all $E>0$.

\Jaindent We consider the part of the orbit from the rightward crossing of the origin at $t=t_{0}$ to the first arrival at the right turning point. Using the conservation law $\dot{q}^{2}=2E-q^{2n}$, we obtain
\begin{align}
	&t-t_{0}=\int_{0}^{q\qty(t)}\dd{s}\,\frac{1}{\sqrt{2E-s^{2n}}}\notag\\
	&\phantom{t-t_{0}}=
	\frac{\qty(2E)^{\qty(1-n)/\qty(2n)}}{2n}
	B_{q\qty(t)^{2n}/\qty(2E)}\qty(\frac{1}{2n},\frac{1}{2}).
	\label{eq:power-law-quadrature}
\end{align}
Here, $B_{z}\qty(a,b)=\int_{0}^{z}\dd{s}\,s^{a-1}\qty(1-s)^{b-1}$ is the incomplete beta function.
The period is four times the traversal time of this interval. In terms of the complete beta function $B\qty(a,b)=B_{1}\qty(a,b)$, it is
\begin{align}
	T=\frac{2}{n}\qty(2E)^{\qty(1-n)/\qty(2n)}
	B\qty(\frac{1}{2n},\frac{1}{2}).
	\label{eq:power-law-period}
\end{align}
For $n=1$, we recover $T=2\pi$.

\subsection{Pendulum}
\label{subsec:pendulum}
\Jaindent We consider the pendulum defined by
\begin{align}
	w\qty(q)=2\sin\qty(\frac{q}{2}),\qquad
	\ddot{q}=-\sin\qty(q).
	\label{eq:pendulum-model}
\end{align}
For $0<E<2$, the motion is libration about $q=0$, with turning points
$q_{\pm}=\pm2\arcsin\qty(\sqrt{E/2})$.
The pair of signs is $\qty(-,+)$, giving $\nu=-1$.

\Jaindent With $m=E/2$ and $z=t-t_{0}$, the librational solution is
\begin{align}
	&q\qty(t)=2\arcsin\qty[\sqrt{m}\,\operatorname{sn}\qty(z\mid m)],\notag\\
	&p\qty(t)=2\sqrt{m}\,\operatorname{cn}\qty(z\mid m),\notag\\
	&Q\qty(t)=2\sqrt{m}\,
	\exp\qty[-\ii\operatorname{am}\qty(z\mid m)],\notag\\
	&T=4K\qty(m).
	\label{eq:pendulum-libration-solution}
\end{align}
Here, $\operatorname{am}\qty(z\mid m)$ is the Jacobi amplitude. The period diverges as $E\to2-0$. At $E=2$, the motion follows a separatrix orbit that approaches the inverted equilibrium asymptotically.

\Jaindent For $E>2$, the motion is rotational.
After one rotation, taking time $T_{\mathrm{rot}}$, $p$ returns to its initial value. Since $w\qty(q+2\pi)=-w\qty(q)$, we have
\begin{align}
	Q\qty(t+T_{\mathrm{rot}})=Q\qty(t)^{*}.
	\label{eq:pendulum-one-rotation}
\end{align}
The period of $Q$ and the Lax operator is therefore $T_{Q}=2T_{\mathrm{rot}}$, corresponding to two rotations.
With $m=2/E$ and $z=\qty(t-t_{0})/\sqrt{m}$, the solution for positive rotation is
\begin{align}
	&q\qty(t)=2\operatorname{am}\qty(z\mid m),\notag\\
	&p\qty(t)=\frac{2}{\sqrt{m}}\operatorname{dn}\qty(z\mid m),\notag\\
	&Q\qty(t)=\frac{2}{\sqrt{m}}\operatorname{dn}\qty(z\mid m)
	-2\ii\operatorname{sn}\qty(z\mid m),\notag\\
	&T_{\mathrm{rot}}\coloneq2\sqrt{m}K\qty(m),\qquad
	T_{Q}\coloneq4\sqrt{m}K\qty(m).
	\label{eq:pendulum-rotation-solution}
\end{align}
Since $Q$ traverses an arc back and forth within the right half-plane, its winding number over $T_{Q}$ is $\nu=0$.
For negative rotation, $Q$ remains in the left half-plane, again giving $\nu=0$.

\section{Fourier coefficients and edge profiles of the double well}
\label{app:double-well-edge}
\Jaindent For the frequency chain obtained from a periodic orbit of the symmetric double well, we consider a semi-infinite chain that retains the sites $r\ge0$ and removes all couplings across the boundary.
We derive the profile of the zero-energy edge mode for the intrawell dn solution and show that this mode is absent for the cn solution spanning both wells.
We first take the outer turning point in Eq.~\eqref{eq:symmetric-dn-solutions} at $t=0$ and set $\Omega_{0}=2\pi/T$.
For $0<E<1/2$, we define a length in harmonic-index units as
\begin{align}
	&\xi\qty(E)\coloneq
	\frac{K\qty(\dfrac{2\sqrt{2E}}{1+\sqrt{2E}})}
	{\pi K\qty(\dfrac{1-\sqrt{2E}}{1+\sqrt{2E}})}.
	\label{eq:double-well-localization}
\end{align}
As shown below, $\xi$ is the exponential decay length of the amplitude of the dn zero-energy edge mode.

\subsection{Fourier components and the dn edge mode}
\Jaindent We denote the solutions in the right and left wells by $q_{+}\qty(t),q_{-}\qty(t)$, respectively.
The Fourier expansion of the dn solution is
\begin{align}
	&\frac{2q_{+}\qty(t)}{\Omega_{0}}
	=1+2\sum_{s=1}^{\infty}
	\frac{\cos\qty(s\Omega_{0}t)}
	{\cosh\qty[s/\xi\qty(E)]}.
	\label{eq:dw-edge-dn-fourier}
\end{align}
Using $q_{-}=-q_{+}$ and integrating $\dot Q=2\ii qQ$, we obtain
\begin{align}
	&Q_{\pm}\qty(t)=\ii\sqrt{2E}\,\e^{\pm\ii\Omega_{0}t}\notag\\
	&\quad\times\exp\qty[
		\pm2\ii\sum_{s=1}^{\infty}
		\frac{\sin\qty(s\Omega_{0}t)}
		{s\cosh\qty[s/\xi\qty(E)]}].
	\label{eq:dw-edge-dn-Q}
\end{align}
The factor $\e^{\pm\ii\Omega_{0}t}$ gives the winding number $\nu_{\pm}=\pm1$, while the remaining factor generates the higher harmonics due to anharmonicity.
Substitution into Eq.~\eqref{eq:frequency-fourier} gives the hopping amplitudes $h_{s}$.

\Jaindent At zero eigenvalue, the equations for the two sublattices decouple:
\begin{align}
	&\sum_{r'\ge0}h_{r-r'}\psi_{r',B}=0,\qquad
	\sum_{r'\ge0}h_{r'-r}^{*}\psi_{r',A}=0.
	\label{eq:dw-edge-zero-equations}
\end{align}
Choosing the zero mode as an eigenstate of the chiral operator gives a state supported on only one sublattice.
The Wiener--Hopf construction of edge modes in chiral chains separates the coupling function into factors containing positive and negative harmonics~\cite{Jones2023}.
For the left well, applying this construction to Eq.~\eqref{eq:dw-edge-dn-Q} and setting $\psi_{r,A}=0$ gives the solution on sublattice B as a one-sided Fourier series:
\begin{align}
	&\sum_{r=0}^{\infty}\psi_{r,B}\e^{\ii r\Omega_{0}t}\notag\\
	&\quad=\psi_{0,B}\exp\qty[
		\sum_{s=1}^{\infty}
		\frac{\e^{\ii s\Omega_{0}t}}
		{s\cosh\qty[s/\xi\qty(E)]}].
	\label{eq:dw-edge-series}
\end{align}
Indeed, multiplication by $Q_{-}$ gives
\begin{align}
	&Q_{-}\qty(t)\sum_{r=0}^{\infty}\psi_{r,B}\e^{\ii r\Omega_{0}t}\notag\\
	&\quad=\ii\sqrt{2E}\,\psi_{0,B}\e^{-\ii\Omega_{0}t}
	\exp\qty[
		\sum_{s=1}^{\infty}
		\frac{\e^{-\ii s\Omega_{0}t}}
		{s\cosh\qty[s/\xi\qty(E)]}].
	\label{eq:dw-edge-negative-harmonics}
\end{align}
The right-hand side contains only strictly negative harmonics, so every coefficient with $r\ge0$ vanishes and Eq.~\eqref{eq:dw-edge-zero-equations} is satisfied.
For the right well, $Q_{+}^{*}=-Q_{-}$ gives the same profile with A and B interchanged.

\Jaindent Differentiating Eq.~\eqref{eq:dw-edge-series} with respect to $t$ and comparing harmonics gives
\begin{align}
	&r\psi_{r,B}=\sum_{s=1}^{r}
	\frac{\psi_{r-s,B}}{\cosh\qty[s/\xi\qty(E)]},
	\qquad r\ge1.
	\label{eq:dw-edge-recurrence}
\end{align}
The zero mode is obtained by calculating the profile from this recurrence relation and normalizing its squared norm to $1$.

\Jaindent Evaluating the coefficients of Eq.~\eqref{eq:dw-edge-series} for large $r$ gives the amplitude tail
\begin{align}
	&\psi_{r,B}\propto r\exp\qty[-r/\xi\qty(E)].
	\label{eq:dw-edge-tail}
\end{align}
As $E\to+0$, $\xi\to+0$, recovering the harmonic limit in which the mode is localized on a single edge site.
As $E\to1/2-0$, the localization length diverges, and expansion of the complete elliptic integrals gives Eq.~\eqref{eq:double-well-edge-limit}.

\subsection{Critical limit}
\Jaindent For the normalized dn zero-energy edge mode, taking $\xi\to\infty$ in Eq.~\eqref{eq:dw-edge-recurrence} at fixed $r$ gives
\begin{align}
	&\frac{\psi_{r,B}}{\psi_{0,B}}\longrightarrow1,
	\qquad\psi_{r,B}\longrightarrow0.
	\label{eq:dw-edge-fixed-r-limit}
\end{align}
As the zero-energy edge mode spreads over more sites, the probability remaining at each fixed site in the normalized state approaches zero.
The relative squared amplitude $\qty|\psi_{r,B}|^{2}/\qty|\psi_{0,B}|^{2}$ used in Fig.~\ref{fig:nonlinear-hopping-edge} (g) instead approaches $1$, giving the horizontal line in the figure.
This constant sequence has a divergent squared norm, and the normalizable zero-energy edge mode is lost in the separatrix limit.

\subsection{Absence of the cn zero-energy edge mode}
\Jaindent For the cn solution, we shift the time origin in Eq.~\eqref{eq:symmetric-cn-solution} by half a period and take the left turning point at $t=0$.
For $E>1/2$, we set
\begin{align}
	&\xi\qty(E)=
	\frac{2K\qty(\dfrac{1+\sqrt{2E}}{2\sqrt{2E}})}
	{\pi K\qty(\dfrac{\sqrt{2E}-1}{2\sqrt{2E}})}.
	\label{eq:dw-cn-fourier-length}
\end{align}
This length appears in the Fourier expansion of the cn solution.
For this solution, $Q$ can be written as
\begin{align}
	&Q\qty(t)=\ii\sqrt{2E}\exp\qty[-4\ii\sum_{\substack{s\ge1\\s\text{ odd}}}
		\frac{\sin\qty(s\Omega_{0}t)}{s\cosh\qty[s/\xi\qty(E)]}].
	\label{eq:dw-cn-Q-factorization}
\end{align}
The same Wiener--Hopf construction~\cite{Jones2023} splits the exponent into positive- and negative-frequency parts, factorizing $Q$ into two exponential factors.
Each factor and its inverse contain harmonics on the same side only and can therefore be removed successively from the zero-eigenvalue equations of the semi-infinite chain.
The only square-summable solution is then $\psi_{r,A}=\psi_{r,B}=0$, so the cn solution has no zero-energy edge mode.

\begin{figure*}[!t]
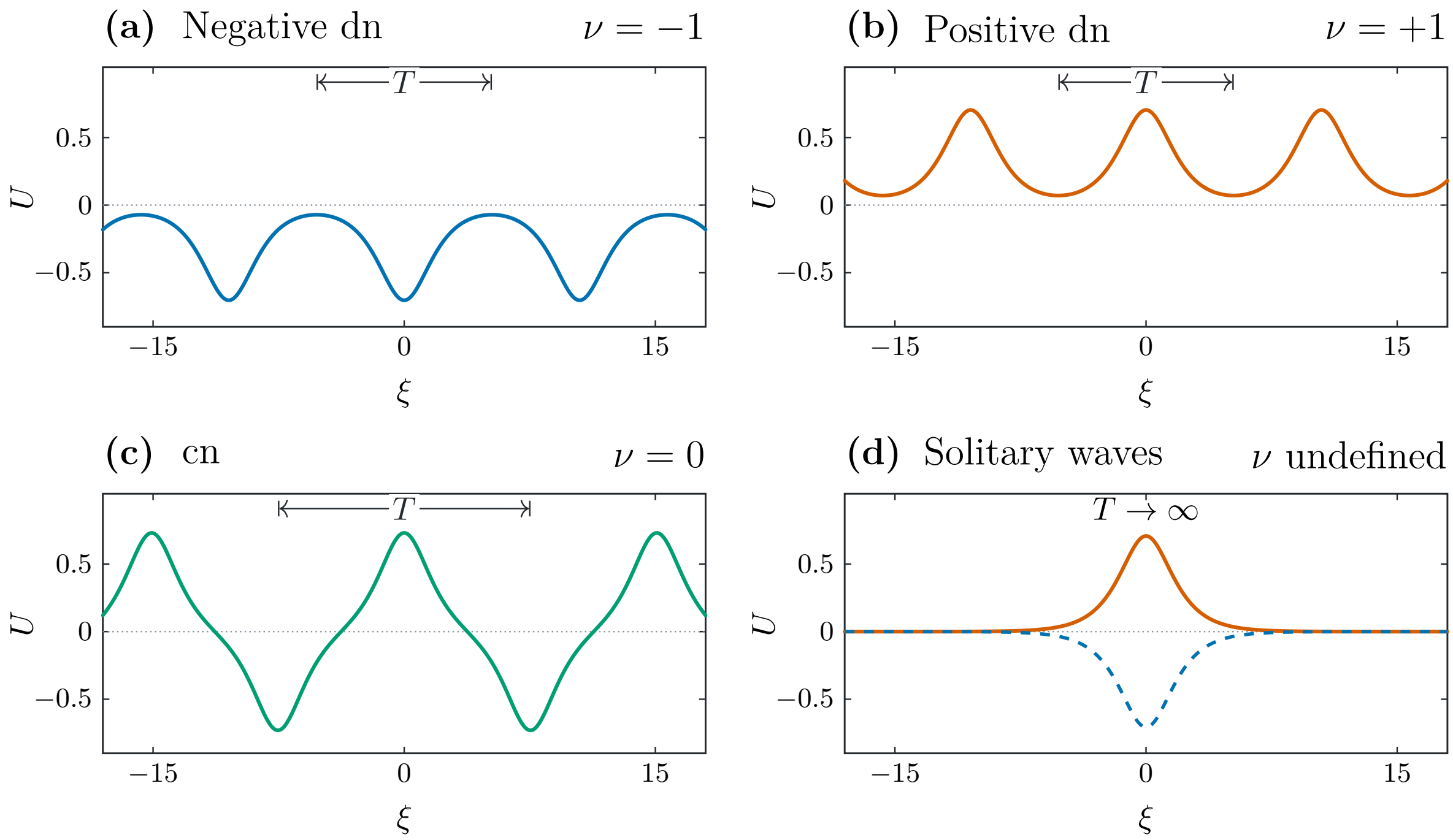

	\centering
	\begingroup%
		\newcommand{\WaveHeadingFont}{\fontsize{12}{14.4}\selectfont}%
		\newcommand{\WavePanelFont}{\fontsize{11.5}{13.8}\selectfont\bfseries}%
		\newcommand{\WaveAnnotationFont}{\fontsize{10}{12}\selectfont}%
		\pgfplotsset{%
			wave/.style={draw=paperorange,paper curve,no marks},%
			negative/.style={draw=paperblue,paper curve,no marks},%
			trivial/.style={draw=paperteal,paper curve,no marks},%
			companion/.style={draw=paperink,line width=1.05pt,dashed,no marks},%
			negative companion/.style={draw=paperblue,line width=1.05pt,dashed,no marks},%
			background/.style={draw=paperguide,densely dotted,line width=.45pt,no marks},%
			every axis/.append style={%
				scale only axis,width=6.5cm,height=2.8cm,axis lines=box,%
				axis line style={paper axis},tick align=inside,%
				tick style={draw=paperink,line width=.45pt},%
				major tick length=2pt,tick label style={font=\PaperLabelFont},%
				label style={font=\WaveAnnotationFont},%
				ylabel style={at={(axis description cs:-.135,.5)},anchor=center},%
				xlabel={$\xi$},ylabel={$U$},%
				scaled ticks=false,clip=false,%
				legend style={font=\PaperSmallFont,draw=none,fill=white,inner sep=2pt}%
			}%
		}%
		\setlength{\fboxsep}{0pt}%
		\setlength{\fboxrule}{0pt}%
		\fbox{%
%
		}%
	\endgroup%
	\caption{Real profiles $U$ common to NLS and mKdV, with $\alpha=1/2$ and $C=0$.
		(a,b) Periodic waves at $E=0.03$. (c) A periodic wave at $E=0.04$. (d) Solitary waves obtained as $E\to1/32\pm0$, approached from below ($-0$) on the dn side and from above ($+0$) on the cn side. The finite-period $\nu$ is undefined for solitary waves.}
	\label{fig:symmetric-traveling-waves}
\end{figure*}

\section{Applications to nonlinear waves}
\label{app:traveling-wave-reductions}
\Jaindent The NLS~\cite{ZakharovShabat1972}, mKdV~\cite{Wadati1973}, and sine--Gordon~\cite{Ablowitz1973SineGordon} equations are integrable nonlinear wave equations that have been solved by the inverse scattering method.
Here, we restrict each equation to traveling-wave solutions and construct the $2\times2$ Lax operator used in this paper from the coordinate and momentum of the resulting system with one degree of freedom.
Evaluating this operator along a periodic wave profile, we examine the winding number of the wave train and the limit in which the profile approaches a solitary wave.
We denote the wave velocity by $v$ and the traveling coordinate by $\xi$ and measure the winding number in the direction of increasing $\xi$.
Subscripts $t,x,\xi$ denote differentiation with respect to the corresponding variables.

\subsection{NLS and mKdV}
\label{subsec:traveling-waves}

\Jaindent We consider the focusing nonlinear Schr\"odinger and mKdV equations,
\begin{align}
	&\ii u_{t}+u_{xx}+2\qty|u|^{2}u=0,\qquad u\in\C,
	\label{eq:focusing-nls}\\
	&u_{t}+6u^{2}u_{x}+u_{xxx}=0,\qquad u\in\R.
	\label{eq:focusing-mkdv}
\end{align}
Setting $\xi=x-vt$, we use a real function $U\qty(\xi)$ and a real constant $\omega$ to write the NLS solution as
\begin{align}
	u\qty(x,t)=\exp\qty{\ii\qty[\frac{vx}{2}+\qty(\omega-\frac{v^{2}}{4})t]}U\qty(\xi).
	\label{eq:nls-traveling-wave}
\end{align}
For mKdV, substitution of $u\qty(x,t)=U\qty(\xi)$ gives
\begin{align*}
	\qty[-vU+2U^{3}+U_{\xi\xi}]_{\xi}=0.
\end{align*}
Choosing the integration constant for mKdV as $C=0$, we reduce both equations to the common profile equation
\begin{align}
	U_{\xi\xi}=\alpha U-2U^{3}.
	\label{eq:nls-mkdv-profile}
\end{align}
Here, $\alpha=\omega$ for NLS and $\alpha=v$ for mKdV. We consider $\alpha\geq0$ below.

\Jaindent Regarding $\xi$ as a time variable, we interpret the profile equation \eqref{eq:nls-mkdv-profile} as the equation of motion of a system with one degree of freedom, with conserved quantity
\begin{align}
	E=\frac{1}{2}\qty(U_{\xi})^{2}
	+\frac{1}{2}\qty(\frac{\alpha}{2}-U^{2})^{2}.
	\label{eq:nls-mkdv-energy}
\end{align}
With $p=U_{\xi}$ and $w\qty(U)=\alpha/2-U^{2}$, this reduces to the Hamiltonian system considered in this paper.
We first take $\alpha>0$, for which the potential is a symmetric double well.
Introducing the variables
\begin{align}
	q=\sqrt{\frac{2}{\alpha}}\,U,\qquad
	\eta=\sqrt{\frac{\alpha}{2}}\,\xi
	\label{eq:nls-mkdv-rescaling}
\end{align}
and denoting differentiation with respect to $\eta$ by a subscript, we obtain
\begin{align}
	&q_{\eta\eta}=2q\qty(1-q^{2}),\notag\\
	&Q\qty(\xi)=U_{\xi}-\ii\qty(\frac{\alpha}{2}-U^{2})\notag\\
	&\phantom{Q\qty(\xi)}=\frac{\alpha}{2}\qty[q_{\eta}-\ii\qty(1-q^{2})].
	\label{eq:nls-mkdv-lax-rescaling}
\end{align}
This change of variables preserves the orientation of $Q$ and its winding number. From Sec.~\ref{subsec:symmetric-double-well}, the negative and positive dn-type periodic waves have $\nu=-1,+1$, respectively, while cn-type periodic waves have $\nu=0$.
Nodeless dn wave trains, sign-alternating cn wave trains, and their solitary-wave limits have previously been studied for NLS~\cite{Carr2000AttractiveNLS}.

\Jaindent For mKdV, this winding number also has a geometric interpretation as the turning number of the tangent to a plane curve~\cite{AncoMaan2026}.
Equation~\eqref{eq:nls-mkdv-profile} gives $Q_{\xi}=2\ii UQ$, so the phase of $Q$ is the tangent angle of a curve whose curvature is $2U$, with $\xi$ as its arc-length parameter.
The values $\nu=\pm1$ for dn waves and $\nu=0$ for cn waves agree with the turning numbers of these curves.
The frequency-chain construction assigns this geometric integer to the chiral band model generated by the periodic wave.

\Jaindent Explicit periodic waves are obtained by replacing the energy in Eqs.~\eqref{eq:symmetric-dn-solutions} and \eqref{eq:symmetric-cn-solution} by $4E/\alpha^{2}$ and $t$ by $\eta$, and setting $U=\sqrt{\alpha/2}\,q$.
The period measured in $\xi$ is $\sqrt{2/\alpha}$ times the period in those equations.
For $\alpha=0$, we have $U_{\xi\xi}=-2U^{3}$ and $w\qty(U)=-U^{2}$; the periodic orbits are the same as those of the $n=2$ power-law model and give $\nu=0$.

\Jaindent The NLS winding number is determined by the real profile $U$.
A constant phase shift of $\pi$ in the complex field gives $U\mapsto-U$ and reverses the sign of $\nu$, while preserving the distinction between the dn and cn types based on $\qty|\nu|$.
If $T$ is the period of $U$, the complex field satisfies $u\qty(x+T,t)=\exp\qty(\ii vT/2)u\qty(x,t)$.

\Jaindent For $\alpha>0$, as $E\to\alpha^{2}/8\pm0$, the pulse spacing grows, and we obtain the solitary wave on the separatrix,
\begin{align}
	U_{\mathrm{sol}}\qty(\xi)=\pm\sqrt{\alpha}\,
	\operatorname{sech}\qty[\sqrt{\alpha}\qty(\xi-\xi_{0})].
	\label{eq:nls-mkdv-solitary-wave}
\end{align}
As shown in Fig.~\ref{fig:symmetric-traveling-waves}, the dn train of pulses with the same sign and the cn train of alternating-sign pulses approach the same solitary wave when individual pulses are considered.
The finite-period winding number distinguishes the wave trains as a whole and becomes undefined in the solitary-wave limit, where the period diverges.

\begin{figure*}[!t]
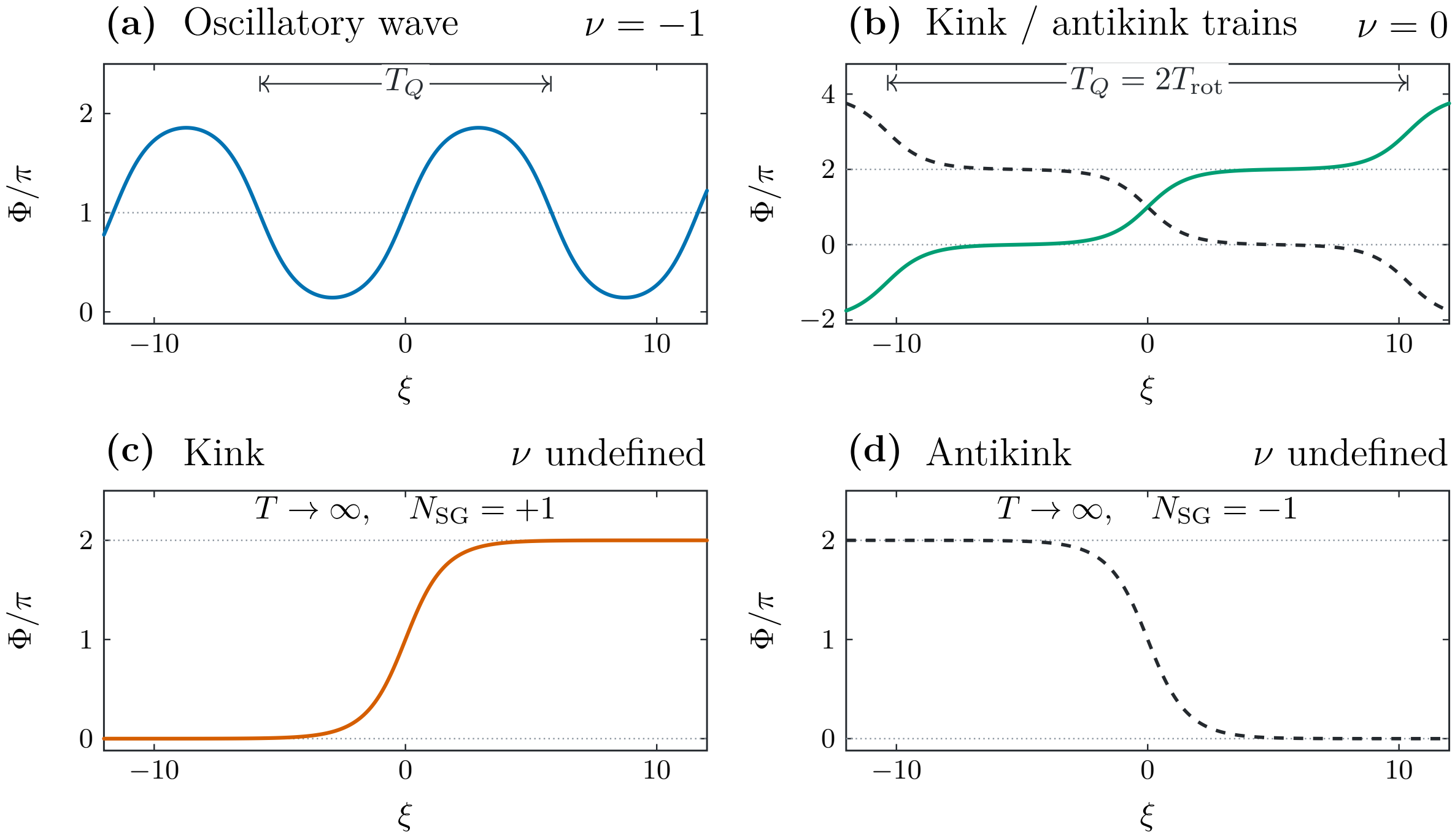

	\centering
	\begingroup%
		\newcommand{\WaveHeadingFont}{\fontsize{12}{14.4}\selectfont}%
		\newcommand{\WavePanelFont}{\fontsize{11.5}{13.8}\selectfont\bfseries}%
		\newcommand{\WaveAnnotationFont}{\fontsize{10}{12}\selectfont}%
		\pgfplotsset{%
			wave/.style={draw=paperorange,paper curve,no marks},%
			negative/.style={draw=paperblue,paper curve,no marks},%
			trivial/.style={draw=paperteal,paper curve,no marks},%
			companion/.style={draw=paperink,line width=1.05pt,dashed,no marks},%
			negative companion/.style={draw=paperblue,line width=1.05pt,dashed,no marks},%
			background/.style={draw=paperguide,densely dotted,line width=.45pt,no marks},%
			every axis/.append style={%
				scale only axis,width=6.5cm,height=2.8cm,axis lines=box,%
				axis line style={paper axis},tick align=inside,%
				tick style={draw=paperink,line width=.45pt},%
				major tick length=2pt,tick label style={font=\PaperLabelFont},%
				label style={font=\WaveAnnotationFont},%
				ylabel style={at={(axis description cs:-.135,.5)},anchor=center},%
				xlabel={$\xi$},ylabel={$U$},%
				scaled ticks=false,clip=false,%
				legend style={font=\PaperSmallFont,draw=none,fill=white,inner sep=2pt}%
			}%
		}%
		\setlength{\fboxsep}{0pt}%
		\setlength{\fboxrule}{0pt}%
		\fbox{%
%
		}%
	\endgroup%
	\caption{Sine--Gordon profiles $\Phi$. (a) $E=1.9$. (b) $E=2.001$, with solid and dashed lines denoting kink and antikink trains, respectively. The horizontal bars indicate one period of $Q$.
		(c,d) Solitary waves obtained as $E\to2\pm0$, approached from below ($-0$) on the librational side and from above ($+0$) on the rotational side. Here, $\nu$ is undefined, and $N_{\mathrm{SG}}$ denotes the topological charge of the field.}
	\label{fig:sine-gordon-traveling-waves}
\end{figure*}

\subsection{Sine--Gordon}

\Jaindent For the sine--Gordon equation,
\begin{align}
	u_{tt}-u_{xx}+\sin\qty(u)=0,\qquad u\in\mathbb{R},
	\label{eq:sine-gordon}
\end{align}
substituting $u\qty(x,t)=\Phi\qty(\xi)$ with $\xi=\qty(x-vt)/\sqrt{1-v^{2}}$ and $\qty|v|<1$ gives $\Phi_{\xi\xi}=\sin\qty(\Phi)$.
Setting $q=\Phi-\pi$, we obtain
\begin{align}
	q_{\xi\xi}=-\sin\qty(q),\qquad
	E=\frac{1}{2}\qty(q_{\xi})^{2}+2\sin\qty(\frac{q}{2})^{2}.
	\label{eq:sine-gordon-reduction}
\end{align}
Replacing $t$ by $\xi$ in Eqs.~\eqref{eq:pendulum-libration-solution} and \eqref{eq:pendulum-rotation-solution}, we obtain periodic oscillatory waves from pendulum libration and kink or antikink trains from rotation. In the latter case, the field changes by $\pm2\pi$ per rotation.
Oscillatory waves have $\nu=-1$, whereas kink and antikink trains have $\nu=0$ over the period of two rotations required for $Q$ to close (Fig.~\ref{fig:sine-gordon-traveling-waves} (a,b)).

\Jaindent As the separatrix is approached, the period diverges and individual kinks become isolated.
Taking $m\to1-0$ on the librational side, we obtain, near a single rising front,
\begin{align}
	&\Phi_{\mathrm{kink}}\qty(\xi)=4\arctan\qty[\exp\qty(\xi-\xi_{0})],\notag\\
	&Q\qty(\xi)=2\qty[\operatorname{sech}\qty(\xi-\xi_{0})
		-\ii\tanh\qty(\xi-\xi_{0})].
	\label{eq:sine-gordon-kink}
\end{align}
Since $Q\to\pm2\ii$ as $\xi\to\mp\infty$, the curve traced by $Q$ is an open semicircle, and the finite-period winding number cannot be defined.
The topological charge of the field, on the other hand, is
\begin{align}
	N_{\mathrm{SG}}=
	\frac{\Phi_{\mathrm{kink}}\qty(+\infty)-\Phi_{\mathrm{kink}}\qty(-\infty)}{2\pi}=1
	\label{eq:sine-gordon-kink-charge}
\end{align}
and is determined by the field values at the two ends.
For the antikink $\Phi_{\mathrm{antikink}}\qty(\xi)=2\pi-\Phi_{\mathrm{kink}}\qty(\xi)$, we have $N_{\mathrm{SG}}=-1$.
The isolated kink and antikink are shown in Fig.~\ref{fig:sine-gordon-traveling-waves} (c,d).

\clearpage
\renewcommand{\bibsection}{

\end{document}